\documentclass[a4paper,11pt]{article}
\usepackage{jheppub}

\usepackage[T1]{fontenc} 
\usepackage{amsmath,epsf,amssymb,latexsym,amsthm,setspace,bbm,longtable}

\usepackage[matrix,arrow,curve,arc]{xy}
\usepackage{graphics,makecell,diagbox}
\usepackage{array,multirow,multicol}
\usepackage{caption,booktabs,subcaption,lipsum}
\usepackage{etoolbox}
\usepackage{titlesec}
\usepackage{cleveref}
\usepackage{float}
\usepackage{lmodern}
\usepackage{subcaption}
\usepackage{xcolor}
\usepackage{pgfplots}
\usepackage{tikz}
\usepackage{placeins}

\usetikzlibrary{arrows, automata, positioning, quotes,shapes}
\pgfdeclarelayer{background,main}
\pgfdeclarelayer{edgelayer}
\pgfdeclarelayer{nodelayer}
\pgfsetlayers{edgelayer,nodelayer,main}
\usetikzlibrary{decorations.markings}

\tikzstyle{none}=[inner sep=0pt]
\tikzset{->-/.style={decoration={
  markings,
  mark=at position 0.5 with {\arrow{>}}},postaction={decorate}}}
\tikzset{->>-/.style={decoration={
  markings,
  mark=at position 0.5 with {\arrow{>>}}},postaction={decorate}}}  
 \tikzset{deco/.style={
              decoration={             
                          markings,   
                          mark=at position 0.5 with {\arrow{}
                                      \node[anchor=\pgfdecoratedangle-90] {#1};
                          }
              },
              postaction={decorate}
      }
  }  
\newsavebox\ltmcbox

\titleformat{\paragraph}
{\normalfont\normalsize\bfseries}{\theparagraph}{1em}{}
\titlespacing*{\paragraph}
{0pt}{3.25ex plus 1ex minus .2ex}{1.5ex plus .2ex}
\makeatletter
\def\@fpheader{\relax}
\makeatother

\theoremstyle{definition}

\crefmultiformat{equation}{(#2#1#3)}%
{ and~(#2#1#3)}{, (#2#1#3)}{ and~(#2#1#3)}

\def\C{{\mathbb C}}
\def\F{{\mathbb F}}

\newcommand{\Z}{\mathbb{Z}}

\newcommand\ft{\widetilde{f}}

\renewcommand{\mod}[1]{\text{ (mod }#1)}

\def\e{{\mathfrak{e}}}
\def\f{{\mathfrak{f}}}
\def\g{{\mathfrak{g}}}

\def\Ins<#1>{I${}_{#1}^s$}
\def\Inns<#1>{I${}_{#1}^{ns}$}
\def\InstarNoMon<#1>{I${}_{#1}^{\ast }$}
\def\Instars<#1>{I${}_{#1}^{\ast s}$}
\def\Instarns<#1>{I${}_{#1}^{\ast ns}$}
\def\su{\operatorname{\mathfrak{su}}}
\def\so{\operatorname{\mathfrak{so}}}
\def\sp{\operatorname{\mathfrak{sp}}}
\def\suText<#1>{$\su({#1})$}
\def\soText<#1>{$\so({#1})$}
\def\spText<#1>{$\sp({#1})$}
\def\eText<#1>{$\e_{#1}$}

\def\pmat<#1,#2>{{\begin{pmatrix} {#1} \\ {#2} \end{pmatrix}}}

\makeatletter
\newcommand{\xleftrightarrow}[2][]{\ext@arrow 3359\leftrightarrowfill@{#1}{#2}}
\makeatother

\newcommand{\xdasharrow}[2][->]{
\tikz[baseline=-\the\dimexpr\fontdimen22\textfont2\relax]{
\node[anchor=south,font=\scriptsize, inner ysep=1.5pt,outer xsep=2.2pt](x){#2};
\draw[shorten <=3.4pt,shorten >=3.4pt,dashed,#1](x.south west)--(x.south east);
}
}

\title{Pairing 6D SCFTs}
\author[a,b]{Peter R. Merkx} 
\affiliation[a]{Department of Mathematics and Computer Science, Wesleyan University, Middletown CT, 06459}
\affiliation[b]{Department of Mathematics, University of California, Davis, Davis CA, 95616}
\emailAdd{pmerkx@math.ucdavis.edu}

\abstract{In this note, we discuss families of orbifolds underlying 6D SCFT F-theory models and find a novel pairing structure in the SCFT landscape. Inspection of the rational functions defining models with a common F-theory endpoint leads us to naturally to pair them and find compatible extended groupings matching endpoint collections recently characterized in correspondence with homomorphisms of the ADE subgroups of $SU(2)$ into $E_8.$ We confirm this proposed pairing closely links the proposed SCFT family pairs via explicit computation of gauge algebras. We find these typically pair precisely by a fixed additional gauge summand. The underlying $\mathbb{C}^2$ orbifold pairing is distinct from the lattice/overlattice orbifold duality which lacks closure on the set of SCFT endpoints. The previously established partial order on endpoints is respected by this pairing as is the distinguished role of certain theories allowing M5-brane fraction reassembly which appear here as self-dual endpoints. This duality manifests in the known tower structure of endpoints to a mirror in a tower below which we show exists naturally as an infinite chain of endpoints extrapolated to negative valuations of the rational functions defining endpoints. We also detail a related simple combinatorial prescription for all rational functions defining endpoint families.}

\begin{document} 
\maketitle
\flushbottom

\begin{section}{Introduction}
Six dimensional superconformal field theories (6D SCFTs) are among the most interesting constructs in high-energy theoretical physics. Global structure in the landscape of these theories continues to be found, as do relationships of this structure with that of the string landscape more generally. In particular, which of these theories compatibly couple with gravity, how their further compactifications relate to the 4D SCFT landscape, and which 6D SCFT renormalization group (RG) flows exist remain to be fully understood.

Over twenty years have passed since the time when questions concerning the existence of 6D SCFTs were put to rest by a variety of constructions of Witten~\cite{Witten:1995zh} and in turn Seiberg~\cite{Seiberg:1996vs}, long after work of Nahm had argued that these theories could exist but would be non-Lagrangian~\cite{Nahm:1977tg}. Relying heavily on tools from F-theory~\cite{Vafa:1996xn,FCY1,FCY2}, dramatic progress classifying these theories has been made in the last several years~\cite{kumar2009bound,kumar2010global,bases,classifyingSCFTs,atomic}. Subsequent refinements to these classification results and characterizations of their global symmetries have followed~\cite{BertoliniGlobal1,gaugeless,gsMerkx}, as have related steps towards classifying 6D RG-flows~\cite{Heckman:2015ola,heckman20166d} and 6D SCFTs involving frozen singularities~\cite{frozen,frozenFTheory,frozenSolo}.

In this note, we focus on global structure emerging in the 6D SCFT landscape at large. In particular, we study relations among the so-called ``endpoint'' families of 6D SCFTs as classified in~\cite{classifyingSCFTs} and endowed with a beautiful unified description as a single tower of theories in~\cite{4DFtheory}. Our main result is that we find a pairing of these endpoints that arises naturally. We will refer to this as a {\it $T^*$ pairing} since it manifests as a transposition in a lattice of endpoints we obtain by extending to $\Z$ the $\mathbb{N}$ valuations of the finitely many rational functions known to define all linear endpoints.

Perhaps most importantly, gauge algebras in $T^*$ paired endpoints typically form precise matchings differing by a uniform gauge summand. Endpoints in paired families contain a set of gauge theories that typically biject via addition of a fixed gauge summand (for example as in Table~\ref{t:TPairAndRanks}). This $T^*$ pairing distinguishes itself as corresponding to a below-ground reflection of the established skyscraper description of endpoints. The known $\mathbb{N}$ valued slices in this tower that capture all linear 6D SCFT endpoints via the aforementioned valuations turn out to have corresponding mirror slices we obtain as negative integer valuations. In carrying out these extrapolations, the $T^*$ duality emerges as a unique, natural reflection swapping positive floors of the tower with those we find exist ``below-ground.''

The associated pairing of $\C^2$ orbifolds differs from the lattice/overlattice $\C^2$ orbifold duality known in the mathematics literature~\cite{Hirzebruch1953,Riemenschneider1974} on resolution of singularities (summarized in~\cite{Reidsurface}). A notable difference is that overlattice duality is not closed on SCFT endpoints. Those endpoints that are fixed under the $T^*$ pairing have appeared previously in work concerning fractional M5-brane probes of partially frozen singularities where they match a class of theories playing a special role with respect to fractional M5 reassembly~\cite{M5Fracs}. 

A larger endpoint grouping structure that we obtain from algebraic invariants of the linear rational functions defining endpoint families is respected by the $T^*$ pairing. This grouping of endpoints matches precisely with that obtained in recent work relating the 6D SCFT landscape with finite ADE subgroup homomorphisms $\Gamma \subset\su(2)\to E_8$~\cite{rgFission,frey20186d}. The related established partial order on endpoints~\cite{heckman2019top} is uniquely respected by this pairing which singles out a distinguished order two operation on the set of endpoints and groupings. Working in reverse from this combinatorial structure allows us to characterize the rational functions that determine SCFT families via a simple set of constraints.

Each 6D SCFT with an F-theory realization having no frozen singularities appears within one of three large groupings determined by the branching of an associated tree of curves. {\it Linear} (or {\it A-type}) {\it endpoints} fall in one of 78 infinite families where there is no branching in these trees. There are 4 infinite {\it D-type endpoint} families where these trees branch, and a handful of exceptional endpoints~\cite{4DFtheory} (corresponding to the Dynkin diagrams of the exceptional Lie groups). Linear endpoints comprise our main focus since one can naturally view the remaining endpoint families as linear endpoint subfamilies for which an additional structure becomes available involving an alternative action on $\C^2.$ We can view a simple structure on all infinite endpoint families as consisting of slices containing finitely many endpoints. We will call these {\it levels} since each endpoint can be described by an unordered pair of integers $1\leq i,j\leq 12$ and a choice of an integer parameter $n$ fixing the slice.

In addition to $T^*$ pairing, we detail relations which rational functions defining endpoint families obey. A pair of constraint equations allows us to provide a prescription for the collection of these functions. The constraints are closely related to the requirement that these functions can be viewed as fractional linear transformations over certain finite fields. A consequence is the replacement of several hundred integer parameters involved in previous endpoint family characterizations with a parsimonious description using only the positive integers $k\leq6.$

At the core of our discussion is an observation motivating $T^*$ pairing that concerns the poles of rational functions determining 6D SCFT endpoints families. To confirm that the pattern obtained which gives rise to $T^*$ pairing is reflected by the SCFTs themselves, we resort to inspection of the realizable gauge algebras in paired endpoints by explicit brute force computation. This analysis reveals that the sets of theories in paired endpoints have a deep structural correspondence and so motivates our further inspection in which we obtain combinatorial endpoint family invariants that reflect known features of the 6D SCFT landscape. It is interesting to note that endpoint family degenerations occurring at small level induce longer orbits via the pairing that endow the landscape with a rich combinatorial structure. Since gauge algebra pairings capture structural similarities between those elliptically fibered CY varieties supported over paired $\C^2/\Gamma$ orbifolds, this pairing and its induced orbits (which we leave implicit here) appear to be of interest to searches for novel substructure in CY threefold moduli space.

The remainder of this note is structured as follows. In Section~\ref{s:review}, we review the $\C^2$ orbifolds underlying F-theory 6D SCFT models and detail canonical orderings of their families compatible with $T^*$ pairing. Enhancement structure respected by $T^*$ pairing is discussed in Section~\ref{s:naturality} where we also briefly contrast $T^*$ pairing with $N^*$ orbifold duality. Section~\ref{s:fracRelations} consists of combinatorial observations concerning the rational functions defining endpoint families including relations obeyed by their integer parameters, family groupings, and steps towards a fraction-intrinsic characterization of the permitted endpoint families. In Appendix~\ref{s:extraps}, we list negative level endpoint extrapolations. In Appendix~\ref{s:contFrac}, we review an alternative condensed description of the 78 rational functions for permitted linear endpoint orbifold actions.  We extend parts of our discussion to D-type endpoints in Appendix~\ref{s:misc} along with a few tables peripheral to our discussion. 
\end{section}

\begin{section}{An SCFT pairing}\label{s:review}
\begin{subsection}{Discrete $U(2)$ gauge field and $\C^2$ orbifold overview}
Recall from~\cite{4DFtheory} that the collection of linear endpoints occurring in infinite families can be expressed as $\alpha A_n \overline{\beta}$ for \begin{align}\label{eq:leads}
\alpha,\overline{\beta}\in H:=\{ 7,6,5,4,3,24,23,223,2223,22223,\emptyset \},
\end{align}
where $\overline{\beta}$ denotes the reverse of a string (e.g.\ for $\alpha=23$, we have $\overline {\alpha} = 32$). Here $\emptyset$ denotes the empty string, and $A_n$ denotes a string $22\cdots2$ of length $n$. We will refer to the strings in $H$ as leads/tails, but will later default to a natural minor variation in this setup. 
Each linear endpoint has an associated orbifold $B\cong C^2/\Gamma$ with $\Gamma\subset U(2)$ a discrete $U(2)$ subgroup with generator
\begin{align}\label{eq:orbAction}
(z_1,z_2)\mapsto (z_1e^{2\pi i \cdot \frac{1}{p}},z_2 e^{2\pi i \cdot \frac{q}{p}}).
\end{align}
We denote the values of $p,q$ which occur for the endpoint $\alpha A_n \beta$ by $p(\alpha A_n \beta),q(\alpha A_n \beta),$ respectively. We similarly define $\frac{p}{q}(\alpha A_n \beta)$ in the obvious way. Values of $p,q$ are given by the Hirzebruch-Jung continued fraction using the endpoint string $\gamma \sim m_1\cdots m_r$ as
\begin{align}\label{eq:contFrac}
\dfrac{p}{q}(m_1,\cdots ,m_r) &= [m_1,\cdots, m_r] 
= m_1- \dfrac{1}{[m_2,\cdots, m_r]} \nonumber \\
 &= m_1 - \cfrac{\text{\scriptsize 1}}{ \text{\scriptsize $m_2 - \cfrac{1}{\overset{\overset{\tiny\ddots
 }{\ }}{\qquad \qquad m_{n-2}}- \cfrac{  \overset{
 }{\ \ 1}}{\text{\tiny $m_{n-1}$}- \frac{1}{m_n }}}$}} \ \underset{\text{\vbox{\hsize=8pt \parindent=0pt \vbox{\ }\vbox{\ }\vbox{\ }\vbox{\ .}}}}{}
\end{align}For clarity, we review the following facts about this setup which is covered in greater depth elsewhere (e.g.~\cite{Reidsurface}). The fraction $p/q$ defines a lattice $L$ given by $\Z^2 + \Z (\frac{1}{p},\frac{q}{p})$ whose convex hull of non-zero points in the first quadrant we refer to as Newton $L.$ Conversely, we can determine uniquely a string of curves from a given rational via a sequence of roundup operations. The invariant monomials in $\C [z_1,z_2]$ for the orbifold determined by $p/q$ are generated by the overlattice dual $N^*(\gamma)\sim\widetilde{m}_1,\cdots \widetilde{m}_{\widetilde{n}}$ corresponding to the fraction $\frac{p}{p-q}.$ More precisely, we can read the generating invariant monomials from the numerators appearing in the overlattice dual endpoint Newton $L$ boundary points when written with minimal common denominator. Note that even when $\gamma$ is a 6D SCFT endpoint, $N^*(\gamma)$ may not be. For example, while $\gamma\sim 3333$ defines an orbifold supporting 6D SCFTs, $N^*(\gamma)\sim 23332$ does not.
\end{subsection}
\begin{subsection}{Defining the pairing}
We begin by inspecting the values of $n$ yielding roots of $p(\alpha A_n \beta)$ in the extrapolated $\frac{p}{q}(\alpha A_n \beta)$ values dating to~\cite{classifyingSCFTs,4DFtheory}. This reveals distinguished matched lead/tail orderings such that all integer valued roots align in a subdiagonal. A unique ordering respects a natural base truncation structure on endpoints as detailed in Section~\ref{s:dets}. Another choice appears in Table~\ref{t:nFracs}. Here, $n$ refers to the level being the number of $-2$ curves rather than the total number of curves $N$ in an endpoint as in~\cite{4DFtheory}. These roots correspond to a breakdown in the orbifold generator action when extrapolating the continued fraction values for $p(\alpha A_n \beta)$ to $n=-2.$
{
\renewcommand*{\arraystretch}{1.8}
     \begin{table}[htbp] 
    \begin{center}
    {      \fontsize{8.2}{10.4}\selectfont\setlength{\tabcolsep}{3pt}
      \begin{tabular}{c|cccccccccccc}
      \diaghead(1,-1){cccccccc}{$\ \alpha$}{$\beta \ $}& 32222 & 3222 & 322 & 32 & 33 & 3 & 42 & 4 & 5 & 6 & 7 & $\emptyset$ \\
      \hline 
 22223  & $ \frac{36 n+96}{30 n+79} $ & $ \frac{30 n+79}{25 n+65} $ & $ \frac{24 n+62}{20 n+51} $ & $ \frac{18 n+45}{15 n+37} $ & $ \frac{30 n+73}{25 n+60} $ & $ \frac{12 n+28}{10 n+23} $ & $ \frac{30 n+67}{25 n+55} $ & $ \frac{18 n+39}{15 n+32} $ & $ \frac{24 n+50}{20 n+41} $ & $ \frac{30 n+61}{25 n+50} $ & $ \frac{36 n+72}{30 n+59} $ & $ \frac{6 n+11}{5 n+9} $\\
 2223  & $ \frac{30 n+79}{24 n+62} $ & $ \frac{25 n+65}{20 n+51} $ & $ \frac{20 n+51}{16 n+40} $ & $ \frac{15 n+37}{12 n+29} $ & $ \frac{25 n+60}{20 n+47} $ & $ \frac{10 n+23}{8 n+18} $ & $ \frac{25 n+55}{20 n+43} $ & $ \frac{15 n+32}{12 n+25} $ & $ \frac{20 n+41}{16 n+32} $ & $ \frac{25 n+50}{20 n+39} $ & $ \frac{30 n+59}{24 n+46} $ & $ \frac{5 n+9}{4 n+7} $\\
 223  & $ \frac{24 n+62}{18 n+45} $ & $ \frac{20 n+51}{15 n+37} $ & $ \frac{16 n+40}{12 n+29} $ & $ \frac{12 n+29}{9 n+21} $ & $ \frac{20 n+47}{15 n+34} $ & $ \frac{8 n+18}{6 n+13} $ & $ \frac{20 n+43}{15 n+31} $ & $ \frac{12 n+25}{9 n+18} $ & $ \frac{16 n+32}{12 n+23} $ & $ \frac{20 n+39}{15 n+28} $ & $ \frac{24 n+46}{18 n+33} $ & $ \frac{4 n+7}{3 n+5} $\\
 23  & $ \frac{18 n+45}{12 n+28} $ & $ \frac{15 n+37}{10 n+23} $ & $ \frac{12 n+29}{8 n+18} $ & $ \frac{9 n+21}{6 n+13} $ & $ \frac{15 n+34}{10 n+21} $ & $ \frac{6 n+13}{4 n+8} $ & $ \frac{15 n+31}{10 n+19} $ & $ \frac{9 n+18}{6 n+11} $ & $ \frac{12 n+23}{8 n+14} $ & $ \frac{15 n+28}{10 n+17} $ & $ \frac{18 n+33}{12 n+20} $ & $ \frac{3 n+5}{2 n+3} $\\
 33  & $ \frac{30 n+73}{12 n+28} $ & $ \frac{25 n+60}{10 n+23} $ & $ \frac{20 n+47}{8 n+18} $ & $ \frac{15 n+34}{6 n+13} $ & $ \frac{25 n+55}{10 n+21} $ & $ \frac{10 n+21}{4 n+8} $ & $ \frac{25 n+50}{10 n+19} $ & $ \frac{15 n+29}{6 n+11} $ & $ \frac{20 n+37}{8 n+14} $ & $ \frac{25 n+45}{10 n+17} $ & $ \frac{30 n+53}{12 n+20} $ & $ \frac{5 n+8}{2 n+3} $\\
 3  & $ \frac{12 n+28}{6 n+11} $ & $ \frac{10 n+23}{5 n+9} $ & $ \frac{8 n+18}{4 n+7} $ & $ \frac{6 n+13}{3 n+5} $ & $ \frac{10 n+21}{5 n+8} $ & $ \frac{4 n+8}{2 n+3} $ & $ \frac{10 n+19}{5 n+7} $ & $ \frac{6 n+11}{3 n+4} $ & $ \frac{8 n+14}{4 n+5} $ & $ \frac{10 n+17}{5 n+6} $ & $ \frac{12 n+20}{6 n+7} $ & $ \frac{2 n+3}{n+1} $\\
 24 & $ \frac{30 n+67}{18 n+39} $ & $ \frac{25 n+55}{15 n+32} $ & $ \frac{20 n+43}{12 n+25} $ & $ \frac{15 n+31}{9 n+18} $ & $ \frac{25 n+50}{15 n+29} $ & $ \frac{10 n+19}{6 n+11} $ & $ \frac{25 n+45}{15 n+26} $ & $ \frac{15 n+26}{9 n+15} $ & $ \frac{20 n+33}{12 n+19} $ & $ \frac{25 n+40}{15 n+23} $ & $ \frac{30 n+47}{18 n+27} $ & $ \frac{5 n+7}{3 n+4} $\\
 4  & $ \frac{18 n+39}{6 n+11} $ & $ \frac{15 n+32}{5 n+9} $ & $ \frac{12 n+25}{4 n+7} $ & $ \frac{9 n+18}{3 n+5} $ & $ \frac{15 n+29}{5 n+8} $ & $ \frac{6 n+11}{2 n+3} $ & $ \frac{15 n+26}{5 n+7} $ & $ \frac{9 n+15}{3 n+4} $ & $ \frac{12 n+19}{4 n+5} $ & $ \frac{15 n+23}{5 n+6} $ & $ \frac{18 n+27}{6 n+7} $ & $ \frac{3 n+4}{n+1} $\\
 5  & $ \frac{24 n+50}{6 n+11} $ & $ \frac{20 n+41}{5 n+9} $ & $ \frac{16 n+32}{4 n+7} $ & $ \frac{12 n+23}{3 n+5} $ & $ \frac{20 n+37}{5 n+8} $ & $ \frac{8 n+14}{2 n+3} $ & $ \frac{20 n+33}{5 n+7} $ & $ \frac{12 n+19}{3 n+4} $ & $ \frac{16 n+24}{4 n+5} $ & $ \frac{20 n+29}{5 n+6} $ & $ \frac{24 n+34}{6 n+7} $ & $ \frac{4 n+5}{n+1} $\\
 6  & $ \frac{30 n+61}{6 n+11} $ & $ \frac{25 n+50}{5 n+9} $ & $ \frac{20 n+39}{4 n+7} $ & $ \frac{15 n+28}{3 n+5} $ & $ \frac{25 n+45}{5 n+8} $ & $ \frac{10 n+17}{2 n+3} $ & $ \frac{25 n+40}{5 n+7} $ & $ \frac{15 n+23}{3 n+4} $ & $ \frac{20 n+29}{4 n+5} $ & $ \frac{25 n+35}{5 n+6} $ & $ \frac{30 n+41}{6 n+7} $ & $ \frac{5 n+6}{n+1} $\\
 7  & $ \frac{36 n+72}{6 n+11} $ & $ \frac{30 n+59}{5 n+9} $ & $ \frac{24 n+46}{4 n+7} $ & $ \frac{18 n+33}{3 n+5} $ & $ \frac{30 n+53}{5 n+8} $ & $ \frac{12 n+20}{2 n+3} $ & $ \frac{30 n+47}{5 n+7} $ & $ \frac{18 n+27}{3 n+4} $ & $ \frac{24 n+34}{4 n+5} $ & $ \frac{30 n+41}{5 n+6} $ & $ \frac{36 n+48}{6 n+7} $ & $ \frac{6 n+7}{n+1} $\\
$\emptyset$  & $ \frac{6 n+11}{6 n+5} $ & $ \frac{5 n+9}{5 n+4} $ & $ \frac{4 n+7}{4 n+3} $ & $ \frac{3 n+5}{3 n+2} $ & $ \frac{5 n+8}{5 n+3} $ & $ \frac{2 n+3}{2 n+1} $ & $ \frac{5 n+7}{5 n+2} $ & $ \frac{3 n+4}{3 n+1} $ & $ \frac{4 n+5}{4 n+1} $ & $ \frac{5 n+6}{5 n+1} $ & $ \frac{6 n+7}{6 n+1} $ & $ \frac{n+1}{n} $\\
       \end{tabular} }
        \caption{Values $\frac{p(n)}{q(n)}$ for continued fraction values of linear endpoints with form $\alpha A_n \beta.$} 
        \label{t:nFracs}
           \end{center}
        \end{table}
 }

This obstruction to negative level extrapolations owes to a root of $p(n)$ (rather than merely non-existence of an SCFT endpoint with specified $p/q$ values). Integer roots of $\frac{p}{q}(\alpha A_n\beta)$ (as in Table~\ref{t:nFracs}) only appear at $n=-2$ with a single exception: the $A_n$ endpoint family singularity appears at $n=-1.$ This turns out to be a feature rather than a bug; that $T^*$ duality can be viewed as a level inversion respecting a canonical partial order of endpoints relies on this fact. 

Computing the roots of $p(n):=p(\alpha A_n \beta)$ with respect to $n$ yields an unexpected link between superficially unrelated families of SCFTs. Reduction of these roots modulo $1$ yields a skew reflection symmetry about the $11\times11$ anti-diagonal, as displayed in Table~\ref{t:reducedPRootsTruncOrder}.

{
\renewcommand*{\arraystretch}{1.8}
     \begin{table}[htbp] 
    \begin{center}
    {      \fontsize{7.2}{9.4}\selectfont\setlength{\tabcolsep}{8pt}
      \begin{tabular}{c|cccccccccccc}
      \diaghead(1,-1){cccccccc}{$\ \alpha$}{$\beta \ $} & 7& 6& 5& 4& 42& 3& 33& 32& 322& 3222& 32222& 2   \\
      \hline 
$ 7 $ & $  -\frac{1}{3} $ & $  -\frac{11}{30} $ & $  -\frac{5}{12} $ & $  \frac{1}{2} $ & $  \frac{13}{30} $ & $  \frac{1}{3} $ & $  \frac{7}{30} $ & $  \frac{1}{6} $ & $  \frac{1}{12} $ & $  \frac{1}{30} $ & $  0 $ & $  -\frac{1}{6} $ \\
$
  6 $ & $  -\frac{11}{30} $ & $  -\frac{2}{5} $ & $  -\frac{9}{20} $ & $  \frac{7}{15} $ & $  \frac{2}{5} $ & $  \frac{3}{10} $ & $  \frac{1}{5} $ & $  \frac{2}{15} $ & $  \frac{1}{20} $ & $  0 $ & $  -\frac{1}{30} $ & $  -\frac{1}{5} $ \\
$
 5 $ & $  -\frac{5}{12} $ & $  -\frac{9}{20} $ & $  \frac{1}{2} $ & $  \frac{5}{12} $ & $  \frac{7}{20} $ & $  \frac{1}{4} $ & $  \frac{3}{20} $ & $  \frac{1}{12} $ & $  0 $ & $  -\frac{1}{20} $ & $  -\frac{1}{12} $ & $  -\frac{1}{4} $ \\
$
 4 $ & $  \frac{1}{2} $ & $  \frac{7}{15} $ & $  \frac{5}{12} $ & $  \frac{1}{3} $ & $  \frac{4}{15} $ & $  \frac{1}{6} $ & $  \frac{1}{15} $ & $  0 $ & $  -\frac{1}{12} $ & $  -\frac{2}{15} $ & $  -\frac{1}{6} $ & $  -\frac{1}{3} $ \\
$
 24 $ & $  \frac{13}{30} $ & $  \frac{2}{5} $ & $  \frac{7}{20} $ & $  \frac{4}{15} $ & $  \frac{1}{5} $ & $  \frac{1}{10} $ & $  0 $ & $  -\frac{1}{15} $ & $  -\frac{3}{20} $ & $  -\frac{1}{5} $ & $  -\frac{7}{30} $ & $  -\frac{2}{5} $ \\
$
 3 $ & $  \frac{1}{3} $ & $  \frac{3}{10} $ & $  \frac{1}{4} $ & $  \frac{1}{6} $ & $  \frac{1}{10} $ & $  0 $ & $  -\frac{1}{10} $ & $  -\frac{1}{6} $ & $  -\frac{1}{4} $ & $  -\frac{3}{10} $ & $  -\frac{1}{3} $ & $  \frac{1}{2} $ \\
$
 33 $ & $  \frac{7}{30} $ & $  \frac{1}{5} $ & $  \frac{3}{20} $ & $  \frac{1}{15} $ & $  0 $ & $  -\frac{1}{10} $ & $  -\frac{1}{5} $ & $  -\frac{4}{15} $ & $  -\frac{7}{20} $ & $  -\frac{2}{5} $ & $  -\frac{13}{30} $ & $  \frac{2}{5} $ \\
$
 23 $ & $  \frac{1}{6} $ & $  \frac{2}{15} $ & $  \frac{1}{12} $ & $  0 $ & $  -\frac{1}{15} $ & $  -\frac{1}{6} $ & $  -\frac{4}{15} $ & $  -\frac{1}{3} $ & $  -\frac{5}{12} $ & $  -\frac{7}{15} $ & $  \frac{1}{2} $ & $  \frac{1}{3} $ \\
$
 223 $ & $  \frac{1}{12} $ & $  \frac{1}{20} $ & $  0 $ & $  -\frac{1}{12} $ & $  -\frac{3}{20} $ & $  -\frac{1}{4} $ & $  -\frac{7}{20} $ & $  -\frac{5}{12} $ & $  \frac{1}{2} $ & $  \frac{9}{20} $ & $  \frac{5}{12} $ & $  \frac{1}{4} $ \\
$
 2223 $ & $  \frac{1}{30} $ & $  0 $ & $  -\frac{1}{20} $ & $  -\frac{2}{15} $ & $  -\frac{1}{5} $ & $  -\frac{3}{10} $ & $  -\frac{2}{5} $ & $  -\frac{7}{15} $ & $  \frac{9}{20} $ & $  \frac{2}{5} $ & $  \frac{11}{30} $ & $  \frac{1}{5} $ \\
$
 22223 $ & $  0 $ & $  -\frac{1}{30} $ & $  -\frac{1}{12} $ & $  -\frac{1}{6} $ & $  -\frac{7}{30} $ & $  -\frac{1}{3} $ & $  -\frac{13}{30} $ & $  \frac{1}{2} $ & $  \frac{5}{12} $ & $  \frac{11}{30} $ & $  \frac{1}{3} $ & $  \frac{1}{6} $ \\
$
 2 $ & $  -\frac{1}{6} $ & $  -\frac{1}{5} $ & $  -\frac{1}{4} $ & $  -\frac{1}{3} $ & $  -\frac{2}{5} $ & $  \frac{1}{2} $ & $  \frac{2}{5} $ & $  \frac{1}{3} $ & $  \frac{1}{4} $ & $  \frac{1}{5} $ & $  \frac{1}{6} $ & $  0 $ \\      
     \end{tabular} }
        \caption{Roots of $p(n)$ reduced modulo 1.} 
        \label{t:reducedPRootsTruncOrder}
           \end{center}
        \end{table}
 }

Lead/tail endpoint families in the final column pair about the $3$-lead row entry. Inspection makes clear that mixed lead pairs are induced by the lead pairing alone. This allows us to summarize the $T^*$ pairing in Figure~\ref{f:pairingTails}. Detailed relation with the natural level inversion structure is discussed in~\ref{s:extraps}. 

Note that the roots of $p(n),q(n)$ modulo 1 are natural level invariants of $f(n).$ They are independent of the $n$-level shift and rescaling $(p,q)\mapsto (a p , a q).$ In other words, for $f_{\alpha,\beta}(n) := f(n)$ and $\ft_k(n):=f(n+k)$ with $k\in \Z_{\geq0}$ we have
\begin{align}
(\ \  f(s) = 0 \quad \text{ and } \quad \ft_k(s')=0 \ \ ) \qquad \implies \qquad  s \equiv s' \mod 1 \ .
\end{align}
These symmetries yield a pairing of endpoints of the form $\alpha A_n \beta$ provided we regard the 6 fixed endpoint orbifold isomorphism classes as paired to themselves (i.e.\ self-dual). We observe that those leads playing a central role upon the introduction of frozen divisors~\cite{frozen,frozenFTheory} also play an exceptional role in $T^*$ pairing, e.g.\ the self-dual leads $2$ and $3$ and the $T^*$ lead pair $23\leftrightarrow4.$
\begin{figure}
  \centering
      \resizebox{0.9\textwidth}{!}{%
\begin{tikzpicture}
[every label/.append style = {font=\tiny},
el/.style = {align=left, font=\tiny},
every  edge/.append style = {draw, shorten > = 12pt,
                             font=\tiny, inner sep=5pt, auto,
                             align=left}]
	\begin{pgfonlayer}{nodelayer}
		\node  (0) at (0, -1) {3};
		\node  (1) at (0, -2) {2};
		\node  (2) at (1, 0) {4};
		\node  (3) at (-1, 0) {32};
		\node  (4) at (2, 1) {5};
		\node  (5) at (-2, 1) {322};
		\node  (6) at (3, 2) {6};
		\node  (7) at (-3, 2) {3222};
		\node  (8) at (-4, 3) {32222};
		\node  (9) at (4, 3) {7};
		\node  (10) at (1, 2) {42};
		\node  (11) at (-1, 2) {33};
		\node (12) at (-6, -2) {1};
		\node (13) at (-6, -1) {2};		
		\node (14) at (-6, 0) {3};				
		\node (15) at (-6, 1) {4};				
		\node (16) at (-6, 2) {5};				
		\node (17) at (-6, 3) {6};				
		\node (18) at (-6, 4) {\tiny$\sqrt{-\det(f(n))}:$};			
		\node (19) at (0, 4.6) {$ \xdasharrow[<->]{\text{$\ \ \ T^*\ \ \ $}} $};		
	\end{pgfonlayer}
	\begin{pgfonlayer}{edgelayer}
		\draw [style={->-}] (1) to (0);
		\draw [style={->-}, bend left=15, looseness=1.00] (0) to (2);
		\draw [style={->-}, bend left=15, looseness=1.00] (2) to (4);
		\draw [style={->-}, bend left=15, looseness=1.00] (4) to (6);
		\draw [style={->-}, bend left, looseness=1.00] (6) to (9);
		\draw [style={->>-}, bend right=15, looseness=1.00] (0) to (3);
		\draw [style={->>-}, bend right=15, looseness=1.00] (3) to (5);
		\draw [style={->>-}, bend right, looseness=1.00] (5) to (7);
		\draw [style={->>-}, bend right=15, looseness=1.00] (7) to (8);
		\draw [style={->-}, bend right=15, looseness=0.75] (3) to (11);
		\draw [style={->>-}, bend left=15, looseness=0.75] (2) to (10);
		\draw [style={deco={\ },dashed}, bend left=15, looseness=1.00] (5) to (4);
		\draw [style={deco={\ },dashed}, bend left=15, looseness=1.0] (8) to (9);
		\draw [style={deco={\ },dashed}, bend left=15, looseness=1.00] (7) to (6);
		\draw [style={deco={\ },dashed}, bend left=15, looseness=0.6] (11) to (10);
		\draw [style={deco={\ },dashed}, bend left=15, looseness=1.00] (3) to (2);
		\draw [style={->-,dotted}, bend left=15, looseness=1.00] (9) to (6);
		\draw [style={->-,dotted}, bend left=15, looseness=1.00] (6) to (4);
		\draw [style={->-,dotted}, bend left=15, looseness=1.00] (4) to (2);
		\draw [style={->-,dotted}, bend right=15, looseness=1.00] (2) to (10);
		\draw [style={->-,dotted}, bend right=15, looseness=0.45] (10) to (0);
		\draw [style={->-,dotted}, bend right=15, looseness=0.45] (0) to (11);
		\draw [style={->-,dotted}, bend right=15, looseness=1.00] (11) to (3);
		\draw [style={->-,dotted}, bend left=15, looseness=1.00] (3) to (5);
		\draw [style={->-,dotted}, bend left=15, looseness=1.00] (5) to (7);
		\draw [style={->-,dotted}, bend left=15, looseness=1.00] (7) to (8);
		\draw [style={->-,dotted}, bend right=15, looseness=1.00] (8) to (1);
		\draw [style={->-,dotted}, bend right=15, looseness=1.00] (1) to (9);
	\end{pgfonlayer}
\end{tikzpicture}
      }%
\caption{$T^*$ lead relations.  Nearest canonical base truncations are indicated by dotted arrows, $T^*$ pairings by dashed arrows. Single solid arrows indicate a transition obtained by blow-up of the rightmost curve followed by truncation of the resulting $-1$ curve. Double arrows indicate a transition by inclusion in a longer base with a rightmost $-1$ curve followed by blow up of this curve and truncation of the resulting $-1$ curve.}
\label{f:pairingTails}
\end{figure}
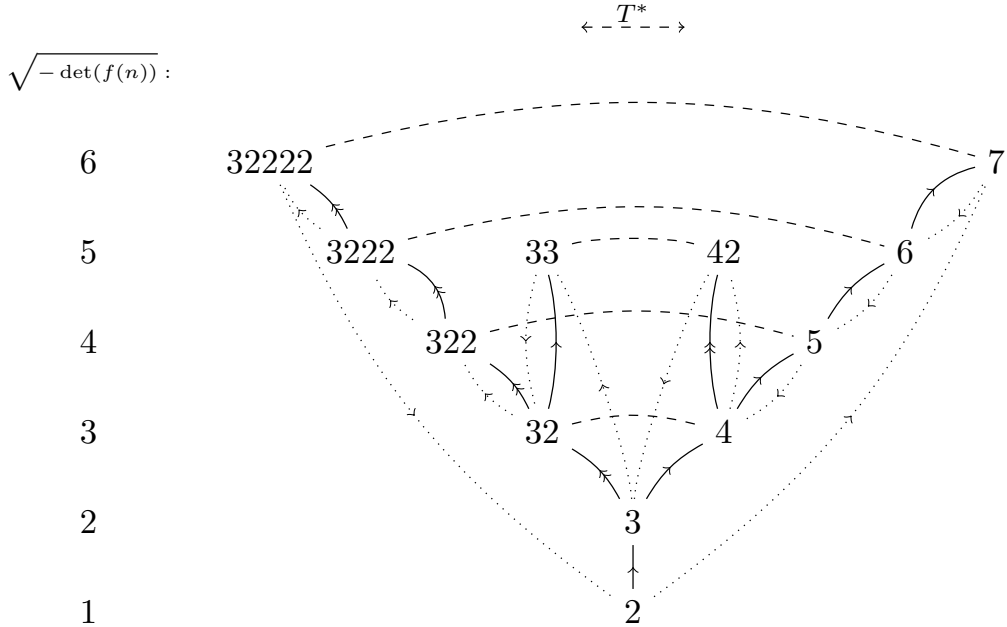%

Each of the rational functions defining an endpoint family takes the form 
\begin{align}\label{eq:genFrac}
f(n) = \dfrac{k_u n + r_u}{k_l n + r_l} \ .
\end{align}
We define the {\it fraction determinant} of $f(n)$ as
\begin{align}\label{eq:fracDet}
\det(f(n)) :=  (k_u\cdot r_l - k_l \cdot r_u)\ , 
\end{align}
noting the invariance under shifts $n\mapsto \widetilde{n}:= n+k$ inducing $f(n)\mapsto \widetilde{f(\widetilde{n})}$ with $f(n),\widetilde{f(\widetilde{n})}$ yielding equivalent sequences in $\mathbb{Q}.$ The value of $\det(f(n))$ is constant on the columns of Table~\ref{t:nFracs} with values appearing in Table~\ref{t:fracDet}. Note that $-\det(f(n))\in \{k^2: 1\leq k\leq 6\}$ with $f(n)$ obeying
\begin{align}\label{eq:fracDetRelation}
-\det(f(n)) = \gcd(k_u,k_l)^2 \ ,
\end{align}
hence requiring that $\gcd(k_u,k_l)$ is also $\alpha$ invariant.
{\renewcommand*{\arraystretch}{1.8}
     \begin{table}[htbp] 
    \begin{center}
    {      \fontsize{8.2}{10.4}\selectfont\setlength{\tabcolsep}{8pt}
      \begin{tabular}{c|ccccccccccccc}
      $\beta:$ &  32222 & 3222 & 4 & 33 & 322 & 3 & 5 & 42 & 32 & 6 & 7 & $\emptyset$ \\
      \hline 
      $-\det(f(\alpha A_n\beta))$: & $6^2$ & $5^2$ & $3^2$ & $5^2$ & $4^2$ & $2^2$ & $4^2$ & $5^2$ & $3^2$ & $5^2$ & $6^2$ & $1$
     \end{tabular} }
        \caption{The ($\alpha$ independent) values of $-\det(f(\alpha A_n\beta)).$} 
        \label{t:fracDet}
           \end{center}
        \end{table}
 }

Note that we can separately reorder the leads and tails distinctly to express $T^*$ pairing for leads on the same footing in a grid as other endpoints. We utilize this ordering to make apparent that $T^*$ pairing respects the block diagonal grouping apparent in Table~\ref{t:intDets} of the determinant action $\det:\Gamma\to U(1)$ originally observed in~\cite{4DFtheory} that we discuss in Section~\ref{s:dets}.

\end{subsection}

\end{section}

\begin{section}{$T^*$ paired gauge algebras}\label{s:naturality}

We now turn to the relationship between gauge algebras within $T^*$ paired endpoints. We restrict our work to non-branching bases in each endpoint pair for computational ease. We compute the gauge algebras on these bases in $T^*$ paired endpoints for a fixed $n$ and find that endpoint pairs often exhibit a pattern of degeneration towards matching at sufficiently large $n.$

We will refer to the number of gauge algebras supported on the linear quivers in an endpoint $\alpha$ as $N_a(\alpha).$ As we move up the tower of levels, endpoint pairs $\alpha\leftrightarrow \beta$ have $N_a(\alpha),N_a(\beta)$ that generally stabilize to achieve near, and typically exact, matching. For example, consider the $T^*$ paired family, namely $22223A_n\leftrightarrow A_n 7.$ Values of $N_a$ in this case appear in Table~\ref{t:countStabilization}. They converge to match exactly for $4\leq n \leq 10$ (and seemingly for all $n\geq 11$). The structure of the algebras on each side of this pairing at $n=10$ is identical up to a uniform addition of a $A_1^2\oplus \f_4 \oplus \g_2^2$ gauge summand as shown in Table~\ref{t:n10PairsGaugeCounts}.

{\small\setlength{\tabcolsep}{0pt}
     \begin{table}[htbp] \footnotesize
    \begin{center}
      \begin{tabular}{l c rl} & $\qquad \quad $&  \multicolumn{2}{c}{ \# Gauge algs. on} \\
   $n$: &  & $\ \ $ $\{22223 A_n$-, & -$A_n 7\}$: \\ 
\hline
1 & & $\{193 ,$ & $30\}$ \\
2 & & $\{128 ,$ & $35\}$ \\
3 & & $\{71 ,$ & $37\}$ \\
\hline
4 & & $\{38 ,$ & $38\}$ \\  
5 & & $\{38 ,$ & $38\}$ \\  
6 & & $\{38 ,$ & $38\}$ \\  
7 & & $\{38 ,$ & $38\}$ \\  
8 & & $\{38 ,$ & $38\}$ \\  
9 & & $\{38 ,$ & $38\}$ \\  
10  & & $\{38 ,$ & $38\}$ \\
	\end{tabular} 
        \caption{Number of gauge algebras on linear bases in the endpoint pairs $22223 A_n\leftrightarrow A_n 7$ for various $n$ values. For fixed $n\geq 4,$ the $T^*$ paired gauge algebra lists differ by a $A_1^2\oplus \f_4\oplus \g_2^2$ summand.} 
        \label{t:countStabilization}
           \end{center}
        \end{table}
 }
 
A second case exhibiting similar behavior appears in Table~\ref{t:TPairAndRanks3} where rank pairing is instead due to uniform addition of $\f_4$ summand. Though frequent, this type of uniform difference pairing does not hold in all cases. Paired sets of enhancements do appear to have similar structure, but can be more involved. For example, Table~\ref{t:TPairAndRanks2} illustrates an additional gauge summand is not always responsible for enhancement count matching.

{\small\setlength{\tabcolsep}{5pt}
     \begin{table}[htbp] \footnotesize
    \begin{center}
      \begin{tabular}{rr rl} Gauge Alg. $\g_a$ on $\alpha \sim 3 A_{10} 33 $ & Gauge Alg. $\g_b$ on $T^*{\alpha}\sim 3 A_{10}42$ & $\text{Rank}(\g_a)$ & $\text{Rank}(\g_b)$ \\
\hline 
{\vspace{-0.1in}}\\
$A_{1}^{22} \oplus e_{8}^{11} \oplus f_{4}^{12} \oplus g_{2}^{22}$ & $A_{1}^{22} \oplus e_{8}^{11} \oplus f_{4}^{11} \oplus g_{2}^{22}$ & $202$ & $198$\\
$A_{1}^{22} \oplus A_{2} \oplus e_{6} \oplus e_{8}^{11} \oplus f_{4}^{12} \oplus g_{2}^{22}$ & $A_{1}^{22} \oplus A_{2} \oplus e_{6} \oplus e_{8}^{11} \oplus f_{4}^{11} \oplus g_{2}^{22}$ & $210$ & $206$\\
$A_{1}^{23} \oplus e_{7} \oplus e_{8}^{11} \oplus f_{4}^{12} \oplus g_{2}^{23}$ & $A_{1}^{23} \oplus e_{7} \oplus e_{8}^{11} \oplus f_{4}^{11} \oplus g_{2}^{23}$ & $212$ & $208$\\
$A_{1}^{23} \oplus e_{8}^{12} \oplus f_{4}^{12} \oplus g_{2}^{23}$ & $A_{1}^{23} \oplus e_{8}^{12} \oplus f_{4}^{11} \oplus g_{2}^{23}$ & $213$ & $209$\\
$A_{1}^{24} \oplus e_{8}^{12} \oplus f_{4}^{12} \oplus g_{2}^{23}$ & $A_{1}^{24} \oplus e_{8}^{12} \oplus f_{4}^{11} \oplus g_{2}^{23}$ & $214$ & $210$\\
$A_{1}^{25} \oplus e_{8}^{12} \oplus f_{4}^{12} \oplus g_{2}^{23}$ & $A_{1}^{25} \oplus e_{8}^{12} \oplus f_{4}^{11} \oplus g_{2}^{23}$ & $215$ & $211$\\
$A_{1}^{24} \oplus e_{8}^{12} \oplus f_{4}^{12} \oplus g_{2}^{24}$ & $A_{1}^{24} \oplus e_{8}^{12} \oplus f_{4}^{11} \oplus g_{2}^{24}$ & $216$ & $212$\\
$A_{1}^{24} \oplus A_{2} \oplus e_{8}^{12} \oplus f_{4}^{12} \oplus g_{2}^{23}$ & $A_{1}^{24} \oplus A_{2} \oplus e_{8}^{12} \oplus f_{4}^{11} \oplus g_{2}^{23}$ & $216$ & $212$\\
             \end{tabular} 
        \caption{Gauge algebras and ranks for all linear quiver based theories in each endpoint of the $n=10$ pair $3 A_{10} 33\leftrightarrow 3 A_{10}42.$ Difference by an $\f_4$ summand accounts for the uniform rank 4 differences.} 
        \label{t:TPairAndRanks3}
           \end{center}
        \end{table}
 } 

{\small\setlength{\tabcolsep}{5pt}
     \begin{table}[htbp] \footnotesize
    \begin{center}
      \begin{tabular}{rr rl} Gauge Alg. $\g_a$ on $\alpha \sim A_4 3 A_{10}$ & Gauge Alg. $\g_b$ on $T^*{\alpha}\sim A_{10}7$ & $\text{Rank}(\g_a)$ & $\text{Rank}(\g_b)$ \\
\hline 
{\vspace{-0.1in}}\\
$A_{1}^{16} \oplus e_{8}^{7} \oplus f_{4}^{8} \oplus g_{2}^{16}$ & $A_{1}^{14} \oplus e_{8}^{7} \oplus f_{4}^{7} \oplus g_{2}^{14}$ & $136$ & $126$\\
$A_{1}^{18} \oplus e_{8}^{8} \oplus f_{4}^{9} \oplus g_{2}^{18}$ & $A_{1}^{16} \oplus e_{8}^{8} \oplus f_{4}^{8} \oplus g_{2}^{16}$ & $154$ & $144$\\
$A_{1}^{20} \oplus B_{3} \oplus e_{7} \oplus e_{8}^{8} \oplus f_{4}^{9} \oplus g_{2}^{18}$ & $A_{1}^{18} \oplus B_{3} \oplus e_{7} \oplus e_{8}^{8} \oplus f_{4}^{8} \oplus g_{2}^{16}$ & $166$ & $156$\\
$A_{1}^{19} \oplus e_{8}^{9} \oplus f_{4}^{10} \oplus g_{2}^{20}$ & $A_{1}^{17} \oplus e_{8}^{9} \oplus f_{4}^{9} \oplus g_{2}^{18}$ & $171$ & $161$\\
$A_{1}^{19} \oplus A_{2} \oplus e_{8}^{9} \oplus f_{4}^{10} \oplus g_{2}^{19}$ & $A_{1}^{17} \oplus A_{2} \oplus e_{8}^{9} \oplus f_{4}^{9} \oplus g_{2}^{17}$ & $171$ & $161$\\
$A_{1}^{19} \oplus A_{2}^{2} \oplus e_{6} \oplus e_{8}^{9} \oplus f_{4}^{10} \oplus g_{2}^{19}$ & $A_{1}^{17} \oplus A_{2}^{2} \oplus e_{6} \oplus e_{8}^{9} \oplus f_{4}^{9} \oplus g_{2}^{17}$ & $179$ & $169$\\
$A_{1}^{21} \oplus e_{7} \oplus e_{8}^{9} \oplus f_{4}^{10} \oplus g_{2}^{21}$ & $A_{1}^{19} \oplus e_{7} \oplus e_{8}^{9} \oplus f_{4}^{9} \oplus g_{2}^{19}$ & $182$ & $172$\\
$A_{1}^{21} \oplus B_{3} \oplus e_{7} \oplus e_{8}^{9} \oplus f_{4}^{10} \oplus g_{2}^{20}$ & $A_{1}^{19} \oplus B_{3} \oplus e_{7} \oplus e_{8}^{9} \oplus f_{4}^{9} \oplus g_{2}^{18}$ & $183$ & $173$\\
$A_{1}^{21} \oplus e_{8}^{10} \oplus f_{4}^{10} \oplus g_{2}^{21}$ & $A_{1}^{19} \oplus e_{8}^{10} \oplus f_{4}^{9} \oplus g_{2}^{19}$ & $183$ & $173$\\
$A_{1}^{21} \oplus e_{8}^{10} \oplus f_{4}^{11} \oplus g_{2}^{21}$ & $A_{1}^{19} \oplus e_{8}^{10} \oplus f_{4}^{10} \oplus g_{2}^{19}$ & $187$ & $177$\\
$A_{1}^{21} \oplus D_{4} \oplus e_{8}^{10} \oplus f_{4}^{10} \oplus g_{2}^{21}$ & $A_{1}^{19} \oplus D_{4} \oplus e_{8}^{10} \oplus f_{4}^{9} \oplus g_{2}^{19}$ & $187$ & $177$\\
$A_{1}^{21} \oplus D_{4}^{2} \oplus e_{8}^{10} \oplus f_{4}^{10} \oplus g_{2}^{21}$ & $A_{1}^{19} \oplus D_{4}^{2} \oplus e_{8}^{10} \oplus f_{4}^{9} \oplus g_{2}^{19}$ & $191$ & $181$\\
$A_{1}^{21} \oplus e_{8}^{10} \oplus f_{4}^{12} \oplus g_{2}^{22}$ & $A_{1}^{19} \oplus e_{8}^{10} \oplus f_{4}^{11} \oplus g_{2}^{20}$ & $193$ & $183$\\
$A_{1}^{21} \oplus A_{2} \oplus e_{8}^{10} \oplus f_{4}^{12} \oplus g_{2}^{21}$ & $A_{1}^{19} \oplus A_{2} \oplus e_{8}^{10} \oplus f_{4}^{11} \oplus g_{2}^{19}$ & $193$ & $183$\\
$A_{1}^{21} \oplus A_{2} \oplus e_{6} \oplus e_{8}^{10} \oplus f_{4}^{11} \oplus g_{2}^{21}$ & $A_{1}^{19} \oplus A_{2} \oplus e_{6} \oplus e_{8}^{10} \oplus f_{4}^{10} \oplus g_{2}^{19}$ & $195$ & $185$\\
$A_{1}^{22} \oplus e_{6} \oplus e_{8}^{10} \oplus f_{4}^{11} \oplus g_{2}^{22}$ & $A_{1}^{20} \oplus e_{6} \oplus e_{8}^{10} \oplus f_{4}^{10} \oplus g_{2}^{20}$ & $196$ & $186$\\
$A_{1}^{22} \oplus e_{7} \oplus e_{8}^{10} \oplus f_{4}^{11} \oplus g_{2}^{22}$ & $A_{1}^{20} \oplus e_{7} \oplus e_{8}^{10} \oplus f_{4}^{10} \oplus g_{2}^{20}$ & $197$ & $187$\\
$A_{1}^{23} \oplus e_{7} \oplus e_{8}^{10} \oplus f_{4}^{11} \oplus g_{2}^{22}$ & $A_{1}^{21} \oplus e_{7} \oplus e_{8}^{10} \oplus f_{4}^{10} \oplus g_{2}^{20}$ & $198$ & $188$\\
$A_{1}^{22} \oplus e_{8}^{11} \oplus f_{4}^{11} \oplus g_{2}^{22}$ & $A_{1}^{20} \oplus e_{8}^{11} \oplus f_{4}^{10} \oplus g_{2}^{20}$ & $198$ & $188$\\
$A_{1}^{23} \oplus e_{8}^{11} \oplus f_{4}^{11} \oplus g_{2}^{22}$ & $A_{1}^{21} \oplus e_{8}^{11} \oplus f_{4}^{10} \oplus g_{2}^{20}$ & $199$ & $189$\\
$A_{1}^{24} \oplus e_{8}^{11} \oplus f_{4}^{11} \oplus g_{2}^{22}$ & $A_{1}^{22} \oplus e_{8}^{11} \oplus f_{4}^{10} \oplus g_{2}^{20}$ & $200$ & $190$\\
$A_{1}^{23} \oplus e_{8}^{11} \oplus f_{4}^{11} \oplus g_{2}^{23}$ & $A_{1}^{21} \oplus e_{8}^{11} \oplus f_{4}^{10} \oplus g_{2}^{21}$ & $201$ & $191$\\
$A_{1}^{23} \oplus A_{2} \oplus e_{8}^{11} \oplus f_{4}^{11} \oplus g_{2}^{22}$ & $A_{1}^{21} \oplus A_{2} \oplus e_{8}^{11} \oplus f_{4}^{10} \oplus g_{2}^{20}$ & $201$ & $191$\\
$A_{1}^{25} \oplus e_{8}^{11} \oplus f_{4}^{11} \oplus g_{2}^{22}$ & $A_{1}^{23} \oplus e_{8}^{11} \oplus f_{4}^{10} \oplus g_{2}^{20}$ & $201$ & $191$\\
$A_{1}^{24} \oplus e_{8}^{11} \oplus f_{4}^{11} \oplus g_{2}^{23}$ & $A_{1}^{22} \oplus e_{8}^{11} \oplus f_{4}^{10} \oplus g_{2}^{21}$ & $202$ & $192$\\
$A_{1}^{24} \oplus A_{2} \oplus e_{8}^{11} \oplus f_{4}^{11} \oplus g_{2}^{22}$ & $A_{1}^{22} \oplus A_{2} \oplus e_{8}^{11} \oplus f_{4}^{10} \oplus g_{2}^{20}$ & $202$ & $192$\\
$A_{1}^{23} \oplus e_{8}^{11} \oplus f_{4}^{11} \oplus g_{2}^{24}$ & $A_{1}^{21} \oplus e_{8}^{11} \oplus f_{4}^{10} \oplus g_{2}^{22}$ & $203$ & $193$\\
$A_{1}^{23} \oplus A_{2} \oplus e_{8}^{11} \oplus f_{4}^{11} \oplus g_{2}^{23}$ & $A_{1}^{21} \oplus A_{2} \oplus e_{8}^{11} \oplus f_{4}^{10} \oplus g_{2}^{21}$ & $203$ & $193$\\
$A_{1}^{23} \oplus A_{2}^{2} \oplus e_{8}^{11} \oplus f_{4}^{11} \oplus g_{2}^{22}$ & $A_{1}^{21} \oplus A_{2}^{2} \oplus e_{8}^{11} \oplus f_{4}^{10} \oplus g_{2}^{20}$ & $203$ & $193$\\
$A_{1}^{23} \oplus B_{3} \oplus e_{8}^{11} \oplus f_{4}^{11} \oplus g_{2}^{23}$ & $A_{1}^{21} \oplus B_{3} \oplus e_{8}^{11} \oplus f_{4}^{10} \oplus g_{2}^{21}$ & $204$ & $194$\\
$A_{1}^{23} \oplus A_{2} \oplus A_{3} \oplus e_{8}^{11} \oplus f_{4}^{11} \oplus g_{2}^{22}$ & $A_{1}^{21} \oplus A_{2} \oplus A_{3} \oplus e_{8}^{11} \oplus f_{4}^{10} \oplus g_{2}^{20}$ & $204$ & $194$\\
$A_{1}^{23} \oplus e_{8}^{11} \oplus f_{4}^{12} \oplus g_{2}^{23}$ & $A_{1}^{21} \oplus e_{8}^{11} \oplus f_{4}^{11} \oplus g_{2}^{21}$ & $205$ & $195$\\
$A_{1}^{23} \oplus D_{4} \oplus e_{8}^{11} \oplus f_{4}^{11} \oplus g_{2}^{23}$ & $A_{1}^{21} \oplus D_{4} \oplus e_{8}^{11} \oplus f_{4}^{10} \oplus g_{2}^{21}$ & $205$ & $195$\\
$A_{1}^{23} \oplus B_{4} \oplus e_{8}^{11} \oplus f_{4}^{11} \oplus g_{2}^{23}$ & $A_{1}^{21} \oplus B_{4} \oplus e_{8}^{11} \oplus f_{4}^{10} \oplus g_{2}^{21}$ & $205$ & $195$\\
$A_{1}^{24} \oplus e_{8}^{11} \oplus f_{4}^{12} \oplus g_{2}^{23}$ & $A_{1}^{22} \oplus e_{8}^{11} \oplus f_{4}^{11} \oplus g_{2}^{21}$ & $206$ & $196$\\
$A_{1}^{23} \oplus e_{8}^{11} \oplus f_{4}^{12} \oplus g_{2}^{24}$ & $A_{1}^{21} \oplus e_{8}^{11} \oplus f_{4}^{11} \oplus g_{2}^{22}$ & $207$ & $197$\\
$A_{1}^{23} \oplus A_{2} \oplus e_{8}^{11} \oplus f_{4}^{12} \oplus g_{2}^{23}$ & $A_{1}^{21} \oplus A_{2} \oplus e_{8}^{11} \oplus f_{4}^{11} \oplus g_{2}^{21}$ & $207$ & $197$\\
$A_{1}^{24} \oplus e_{8}^{11} \oplus f_{4}^{12} \oplus g_{2}^{24}$ & $A_{1}^{22} \oplus e_{8}^{11} \oplus f_{4}^{11} \oplus g_{2}^{22}$ & $208$ & $198$\\
             \end{tabular} 
        \caption{Gauge algebras and ranks for the theories in each endpoint of the $n=10$ pair $ A_4 3 A_{10}\leftrightarrow A_{10}7.$ Difference by an $A_1^2\oplus \f_4\oplus \g_2^2$ summand accounts for the uniform rank 10 differences.} 
        \label{t:TPairAndRanks}
           \end{center}
        \end{table}
 }

The general correspondence appears to be more subtle in two ways. First, a rich paired combinatorial enhancement structure typically holds for small $n$ before a tight correspondence arises. Whether this owes to $T^*$ pairing corresponding to a deeper duality of field theories and/or CY moduli space sectors remains unclear. Evidence against fixed $p(n)$ singularity behavior being responsible for the observed gauge algebra family pairing phenomena is provided by drastically different enhancement counts in non $T^*$ paired families sharing modulo 1 reductions in their roots of $p(n)$. Second, a few branching bases are in some cases allowed. We do not include branching bases in our computations of the permitted gauge enhancements for computational ease at the cost of introducing typically minor paired enhancement count mismatches even for large $n.$ 

Endpoints $\alpha$ with $N_a(\alpha)=\infty$
at level $n=0$ are identified
with two exceptions eliminated provided we pair entire endpoint families across levels. We include an explicit listing of the $T^*$ endpoint pairs in Table~\ref{t:thePairing} where cases permitting infinitely many enhancements are indicated. One of these is the pairing $33A_n\leftrightarrow 24A_n$ which has the endpoint $33$ that can easily be seen to hold infinitely many enhancements (since its minimal resolution, the quiver $414,$ permits infinitely many enhancements) while the endpoint $24$ contains quivers that support only finitely many enhancements. The latter count however becomes infinite when extending this family to $n=-1.$ (Interestingly, $N_a(24)\approx 216$, is perhaps the largest among finite $N_a$ at $n=0.$)\footnote{
To confirm that the endpoints $5,6,322$ and $3222$ have $N_a=\infty$, two approaches are helpful. For $5$ and $6,$ we can compare to the ``long-bases'' Appendix B of~\cite{atomic} (in particular B.22) to find infinitely many distinctly gauged bases blow down to $6$ with $5$ following after decorating by a rightmost $-1$ curve. For $322$ and $3222,$ we can blow up to obtain $(1)4141\cdots$ with $N_a=\infty.$
}

{\small\setlength{\tabcolsep}{5pt}
     \begin{table}[htbp] \footnotesize
    \begin{center}
      \begin{tabular}{rr rl} Gauge Alg. $\g_a$ on $\alpha \sim 4 A_{10} 7 $ & Gauge Alg. $\g_b$ on $T^*{\alpha}\sim A_4 3 A_{10}32$ & $\text{Rank}(\g_a)$ & $\text{Rank}(\g_b)$ \\
\hline 
{\vspace{-0.1in}}\\
$A_{1}^{20} \oplus e_{7} \oplus e_{8}^{10} \oplus f_{4}^{10} \oplus g_{2}^{20}$ & $A_{1}^{23} \oplus e_{8}^{11} \oplus f_{4}^{11} \oplus g_{2}^{23}$ & $187$ & $201$\\
$A_{1}^{20} \oplus e_{8}^{11} \oplus f_{4}^{10} \oplus g_{2}^{20}$ & $A_{1}^{23} \oplus e_{8}^{11} \oplus f_{4}^{12} \oplus g_{2}^{23}$ & $188$ & $205$\\
$A_{1}^{21} \oplus e_{8}^{11} \oplus f_{4}^{10} \oplus g_{2}^{20}$ & $A_{1}^{23} \oplus D_{4} \oplus e_{8}^{11} \oplus f_{4}^{11} \oplus g_{2}^{23}$ & $189$ & $205$\\
             \end{tabular} 
        \caption{Gauge algebras and ranks for the theories in each endpoint of the $n=10$ pair $ 4 A_{10} 7\leftrightarrow A_4 3 A_{10}32$. Here gauge algebra pairs do not differ by a uniform factor.} 
        \label{t:TPairAndRanks2}
           \end{center}
        \end{table}
 }

\begin{subsection}{Gauge algebra counts for all $T^*$ pairs}

To confirm that $T^*$ pairing yields a correspondence between SCFT gauge algebras in $T^*$ paired endpoints more generally, we focus our attention on these pairs at $n=10.$ This value is large enough to avoid endpoint family mixing which may obscure $N_a$ matching and reduce the role of branching bases that we wish to discard for computational simplicity. Note that matching level $n$ pairs appears to be more natural than pairing endpoints with the same total number $N$ of curves for several reasons. First, there is much closer matching of $n$-level pair algebra counts versus those for $N$-level pairs. Second, the correspondence with $N^*$ duality appears to closely resemble $T^*$ pairings at fixed $n.$ Finally, the $n$-level expressions rather than those for $N$-level yield a $p$ singularity structure crucial t o the existence of a natural level-inversion respecting partial order of endpoints induced by truncation of certain distinguished bases.
  
The $N_a$ counts for all $T^*$ pairs at $n=10$ appear as Table~\ref{t:n10PairsGaugeCounts}. We determine $N_a(\alpha)$ values by explicit computation of the permitted gauge algebras using the ancillary code of~\cite{gsMerkx}. This entails first computing all bases in each endpoint and decorating their curves with any globally compatible gauge summands. These results rely in some cases on conjectural existence claims that may also be partly responsible for the minority of cases exhibiting $N_a$ mismatches even at large $n.$ However, since we have only involved linear bases in our SCFT counts, the role of branching bases supporting SCFTs that blow down to a linear endpoint is likely a primary culprit. 

{\small\setlength{\tabcolsep}{0pt}
     \begin{table}[htbp] \footnotesize
    \begin{center}
      \begin{tabular}{rcl rcl crl} \multicolumn{6}{c}{End-pair} & &  \multicolumn{2}{c}{ \# Enhancements} \\
\hline
$\{ 22223$ & $A_{10}$ & $32222,$ & $7$ & $A_{10}$ & $7\}$  & $\qquad$ & $\{1 ,$ & $1 \}$ \\
$\{ 22223$ & $A_{10}$ & $3222,$ & $ 6$ & $A_{10}$ & $7\}$  & $\qquad$ & $\{ 1,$ & $1 \}$ \\
$\{ 22223$ & $A_{10}$ & $33,$ & $24$ & $A_{10}$ & $7\}$  & $\qquad$ & $\{1 ,$ & $1 \}$ \\
$\{ 22223$ & $A_{10}$ & $42,$ & $33$ & $A_{10}$ & $7\}$  & $\qquad$ & $\{1 ,$ & $1 \}$ \\
$\{ 22223$ & $A_{10}$ & $6,$ & $2223$ & $A_{10}$ & $7\}$  & $\qquad$ & $\{1 ,$ & $ 1\}$ \\
$\{ 22223$ & $A_{10}$ & $322,$ & $5$ & $A_{10}$ & $7\}$  & $\qquad$ & $\{2 ,$ & $1 \}$ \\
$\{ 22223$ & $A_{10}$ & $5,$ & $223$ & $A_{10}$ & $7\}$  & $\qquad$ & $\{1 ,$ & $2 \}$ \\
$\{ 22223$ & $A_{10}$ & $4,$ & $23$ & $A_{10}$ & $7\}$  & $\qquad$ & $\{3 ,$ & $3 \}$ \\
$\{ 22223$ & $A_{10}$ & $3,$ & $3$ & $A_{10}$ & $7\}$  & $\qquad$ & $\{8 ,$ & $8 \}$ \\
$\{ 22223$ & $A_{10}$ & $32,$ & $ 4$ & $A_{10}$ & $7\}$  & $\qquad$ & $\{3 ,$ & $3 \}$ \\
$\{ 22223$ & $A_{10}$- & $,$ &  - & $A_{10}$ & $7\}$  & $\qquad$ & $\{38 ,$ & $38\}$ \\
$\{ 6$ & $A_{10}$ & $6,$ & $2223$ & $A_{10}$ & $3222\}$  & $\qquad$ & $\{1 ,$ & $1 \}$ \\
$\{ 6$ & $A_{10}$ & $33,$ & $ 24$ & $A_{10}$ & $3222\}$  & $\qquad$ & $\{1 ,$ & $1\}$ \\
$\{ 6$ & $A_{10}$ & $42,$ & $33$ & $A_{10}$ & $3222\}$  & $\qquad$ & $\{1 ,$ & $1 \}$ \\
$\{ 6$ & $A_{10}$ & $322,$ & $5$ & $A_{10}$ & $3222\}$  & $\qquad$ & $\{2 ,$ & $ 1\}$ \\
$\{ 6$ & $A_{10}$ & $5,$ & $ 223$ & $A_{10}$ & $3222\}$  & $\qquad$ & $\{1 ,$ & $2 \}$ \\
$\{ 6$ & $A_{10}$ & $4,$ & $23$ & $A_{10}$ & $3222\}$  & $\qquad$ & $\{3 ,$ & $ 3\}$ \\
$\{ 6$ & $A_{10}$ & $3,$ & $3$ & $A_{10}$ & $3222\}$  & $\qquad$ & $\{8 ,$ & $8 \}$ \\
$\{ 6$ & $A_{10}$ & $32,$ & $ 4$ & $A_{10}$ & $3222\}$  & $\qquad$ & $\{3 ,$ & $3 \}$ \\
$\{ 6$ & $A_{10}$- & $,$ &  - & $A_{10}$ & $3222\}$  & $\qquad$ & $\{38 ,$ & $38 \}$ \\
$\{ 24$ & $A_{10}$ & $42,$ & $33$ & $A_{10}$ & $33\}$  & $\qquad$ & $\{1 ,$ & $1 \}$ \\
$\{ 24$ & $A_{10}$ & $322,$ & $ 5$ & $A_{10}$ & $33\}$  & $\qquad$ & $\{ 2,$ & $1 \}$ \\
$\{ 24$ & $A_{10}$ & $5,$ & $223$ & $A_{10}$ & $33\}$  & $\qquad$ & $\{ 1,$ & $ 2 \}$ \\
$\{ 24$ & $A_{10}$ & $4,$ & $23$ & $A_{10}$ & $33\}$  & $\qquad$ & $\{3 ,$ & $3 \}$ \\
$\{ 24$ & $A_{10}$ & $3,$ & $ 3$ & $A_{10}$ & $33\}$  & $\qquad$ & $\{8 ,$ & $ 8\}$ \\
$\{ 24$ & $A_{10}$ & $32,$ & $4$ & $A_{10}$ & $33\}$  & $\qquad$ & $\{3 ,$ & $3 \}$ \\
$\{ 24$ & $A_{10}$- & $,$ & -  & $A_{10}$ & $33\}$  & $\qquad$ & $\{38 ,$ & $38\}$ \\
$\{ 5$ & $A_{10}$ & $5,$ & $223$ & $A_{10}$ & $322\}$  & $\qquad$ & $\{2 ,$ & $4 \}$ \\
$\{ 5$ & $A_{10}$ & $4,$ & $ 23$ & $A_{10}$ & $322\}$  & $\qquad$ & $\{ 4,$ & $ 8\}$ \\
\hline
$\{ 5$ & $A_{6}$ & $3,$ & $3$ & $A_{6}$ & $322\}$  & $\qquad$ & $\{12 ,$ & $19 \}$ \\
$\{ 5$ & $A_{6}$ & $32,$ & $ 4$ & $A_{6}$ & $322\}$  & $\qquad$ & $\{ 5,$ & $7 \}$ \\
$\{ 5$ & $A_{6}$- &   & -  & $A_{6}$ & $322\}$  & $\qquad$ & $\{ 64,$ & $95\}$ \\
$\{ 23$ & $A_{6}$ & $32,$ & $4$ & $A_{6}$ & $4\}$  & $\qquad$ & $\{ 10,$ & $ 8\}$ \\
$\{ 23$ & $A_{6}$ & $3,$ & $ 3$ & $A_{6}$ & $4\}$  & $\qquad$ & $\{34 ,$ & $26 \}$ \\
$\{ 23$ & $A_{6}$- & $,$ & - & $A_{6}$ & $4 \}$  & $\qquad$ & $\{168 ,$ & $143 \}$\\                  
             \end{tabular} 
        \caption{Number of gauge algebras arising in paired endpoints. Each non-self-dual endpoint family pair appears once. The first grouping gives values at $n=10,$ while the second is listed for $n=5$ for computational ease.} 
        \label{t:n10PairsGaugeCounts}
           \end{center}
        \end{table}
 }
 
\end{subsection}

\begin{subsection}{More than pairing: larger groupings}\label{s:groupings}

Collections of $T^*$ pairs which have common $N_a$ values are evident. We can determine directly from the rational functions $f(n)$ which endpoints to view as naturally grouped in accordance with these counts having (near) $N_a$ agreement. The values of unordered pairs $\{\det f_{\alpha,\beta}(n),\det(f_{\bar{\beta},\bar{\alpha}}(n))\}$ that we display in Table~\ref{t:intPairs} of Section~\ref{s:fracRelations} suffice to capture and slightly refine these candidate endpoint groupings having similar $N_a$ counts. This appears to suggest that $T^*$ pairing may be one of a more involved collection of operations with orbits consisting of corresponding SCFT structures.

Indeed, these groupings have appeared in recent literature~\cite{rgFission} (cf.\ fixed $\g_L,\g_R$ pairs of Table 1) concerning 6D SCFT RG flows, nilpotent orbits in semisimple Lie groups, and homomorphisms of finite ADE subgroups of $SU(2)$ into $E_8.$ Brief inspection reveals those 6D SCFT endpoints sharing corresponding nilpotent orbits form precisely the same endpoint groupings captured here via fraction determinant pairs.

A matching grouping of endpoints determined by fixed fraction determinant pairs also arise geometrically from the orbifolds $B\cong\C^2/\Gamma.$ The generating invariant monomials $p_1,p_2\cdots, p_k$ of the action by $\Gamma$ give rise to the ring $R\cong\C[p_1,p_2,\cdots,p_k]$~(cf.~\cite{Reidsurface}).
For $n\geq 5,$ all endpoints fall into a unique family. There, it is natural to form the Hasse diagram obtained by inclusion of these rings. Collapsing this graph by identifying all vertices corresponding to endpoints within each single family yields a graph on endpoint families. Provided we carry out this process for those families not involving pure leads/tails for which the situation is more subtle, we find a graph comprised of several connected components, each having automorphisms respecting the aforementioned endpoint family groupings. In other words, the family grouping structure simply characterized here via fraction determinant pairs appears directly via the orbifold geometry. It would be instructive to extend a similar study to graphs formed instead via inclusions between the vector spaces naturally associated to orbifolds as a quotients $R/\sim,$ where $\sim$ arises via the relations between invariant monomials. 
 
\end{subsection}

\begin{subsection}{$T^*$ versus $N^*$ orbifold pairs}\label{s:dualOverlattice}

We now briefly discuss the interplay between $T^*$ and $N^*$ orbifold pairings. The latter pairs the orbifold generator action induced lattice $L$ obtained from a string of curves $\gamma$ and its dual-overlattice $N^*(L)$ from which we can read off the invariant monomials of this action. (For a complete discussion, see~\cite{Reidsurface}). This overlattice in turn corresponds to an orbifold given by a string of curves we denote as $N^*(\gamma)$ for convenience.

In Table~\ref{t:twoPairings}, we display the two pairings together at $n=0.$ Though always distinct, a similarity between $T^*$ and $N^*$ pairs is evident.

\noindent\begin{minipage}{\textwidth}
 \begin{multicols}{2}
 \setbox\ltmcbox\vbox{
 \makeatletter\col@number\@ne
 { \fontsize{6.3}{7.3}\selectfont\setlength{\tabcolsep}{3pt} \begin{longtable}{rccl} $\alpha:$  &  $T^*(\alpha):$ & $N^*(\alpha):$ & $N^*(T^*(\alpha)):$\\
 \hline
  77 & 2222332222 & 22222322222${}^\dagger$ & 636${}^\dagger$ \\
  76 & 222332222 & 2222232222 & 536${}^\dagger$ \\
  75 & 22332222 & 222223222 & 436${}^\dagger$ \\
  74 & 2332222 & 22222322 & 336 \\
  73 & 332222 & 2222232 & 236 \\
  733 & 2432222 & 22222332${}^\dagger$ & 3236${}^\dagger$ \\
  742 & 3332222 & 22222323${}^\dagger$ & 2336${}^\dagger$ \\
  732 & 432222 & 2222233 & 2236 \\
  7322 & 532222 & 2222234${}^\dagger$ & 22236 \\
  73222 & 632222 & 2222235${}^\dagger$ & 222236 \\
  732222 & 732222 & 2222236${}^\dagger$ & 2222236${}^\dagger$ \\
  7 & 32222 & 222222 & 26 \\
  66 & 22233222 & 222232222 & 535${}^\dagger$ \\
  65 & 2233222 & 22223222 & 435${}^\dagger$ \\
  64 & 233222 & 2222322 & 335 \\
  63 & 33222 & 222232 & 235 \\
  633 & 243222 & 2222332 & 3235${}^\dagger$ \\
  642 & 333222 & 2222323 & 2335${}^\dagger$ \\
  632 & 43222 & 222233 & 2235 \\
  6322 & 53222 & 222234 & 22235 \\
  63222 & 63222 & 222235 & 222235 \\
   6 & 3222 & 22222 & 25 \\
  55 & 223322 & 2223222 & 434${}^\dagger$ \\  
   $\vdots$ & $\vdots$ & $\vdots$ \\
       & & \\
   $\alpha:$  &  $T^*(\alpha):$ & $N^*(\alpha):$ & $N^*(T^*(\alpha))$\\    
 \hline  
   $\vdots$ & $\vdots$ & $\vdots$ \\  
  54 & 23322 & 222322 & 334 \\
  53 & 3322 & 22232 & 234 \\
  533 & 24322 & 222332 & 3234${}^\dagger$ \\
  542 & 33322 & 222323 & 2334${}^\dagger$ \\
  532 & 4322 & 22233 & 2234 \\
  5322 & 5322 & 22234 & 22234 \\
  5 & 322 & 2222 & 24 \\
  44 & 2332 & 22322 & 333 \\
  43 & 332 & 2232 & 233 \\
  433 & 2432 & 22332 & 3233 \\
  442 & 3332 & 22323 & 2333 \\
  432 & 432 & 2233 & 2233 \\
  4 & 32 & 222 & 23 \\
  33 & 42 & 232 & 223 \\
  333 & 342 & 2332 & 2323 \\
  3 & 3 & 22 & 22 \\
  3333 & 2442 & 23332${}^\dagger$ & 32323${}^\dagger$ \\
  3342 & 3342 & 23323${}^\dagger$ & 23323${}^\dagger$ \\
        \caption{\footnotesize $T^*$ versus $N^*$ dual-overlattice pairings for level $n=0$ endpoints. Entries marked with $\dagger$ are not valid 6D SCFT endpoints.} 
        \label{t:twoPairings}
   \end{longtable} }%
 \unskip
 \unpenalty
 \unpenalty}
 \unvbox\ltmcbox
 \medskip
 \end{multicols}%
 \end{minipage}%
 
 \begin{subsubsection}{$n\to \infty$ endpoint extrapolations}
Another apparent relation between $T^*,N^*$ orbifold pairings appears when we consider orbifolds appearing in the limit $\alpha_{j,\infty} \leftrightarrow k_{u,\infty}/k_{l,\infty}$ via the Hirzebruch-Jung continued fraction sequence limit $\lim_{n\to\infty}f(n)$. Each such limit gives a valid 6D SCFT endpoint. We collect these in Table~\ref{t:largeNLimits}. For example, $\alpha\sim 33$ has $\alpha A_n$ with a large-$N$ limit matching the continued fraction for $23.$ 

This set of large $n$ limiting ends is closed under the $N^*$ duality which induces a permutation on these endpoints that nearly matches the corresponding $T^*$ duality on lead endpoints $\overline{\alpha}.$ The only exception involves replacement of the orbit $33\overset{T^*}{\leftrightarrow} 42$ with a trivial cycle. In other words, the $N^*$ pairing in the large $n$ limit closely resembles $T^*$ pairing at finite $n$.
 {\renewcommand*{\arraystretch}{1.8}
      \begin{table}[htbp]
     \begin{center}
     {      \fontsize{8.2}{10.4}\selectfont\setlength{\tabcolsep}{8pt}
       \begin{tabular}{c|ccccccccccccc}
       $\overline{\alpha}:$ &  32222 & 3222 & 4 & 33 & 322 & 3 & 5 & 42 & 32 & 6 & 7 & $\emptyset$ \\
       \hline 
       $\alpha A_{\infty}$: & $A_5$ & $ A_4$ & $ A_2$ & $23$ & $ A_3$ & $32$ & 4 & 2 & 3 & 5 & 6 & $\emptyset$\\
       $N^*(\alpha A_{\infty})$: & $6$ & $ 5$ & $ 3$ & $32$ & $ 4$ & $23$ & $A_3$ & 2 & $A_2$ & $A_4$ & $A_5$ & $\emptyset$      
      \end{tabular} }
         \caption{Ends extrapolated from $\alpha A_n$ as $n\to \infty$ and their overlattice duals.} 
         \label{t:largeNLimitExtraps}
            \end{center}
         \end{table}
  }
 
 \end{subsubsection}

\end{subsection}
\end{section}

\begin{section}{On 6D SCFT endpoint combinatorics}\label{s:fracRelations}

In this section we provide a few observations concerning rules governing the collection of 6D SCFT endpoints with a focus on the paired rational functions $f(n)$ in terms of their integer ingredients and overall structure. We begin by working towards a simple description of the permitted linear 6D SCFT endpoint continued fraction formulas. We uncover an identity holding all $f(n)$ defining endpoint families. This turns out to be sufficient to characterize those families purely via the fraction determinant values. We then briefly discuss how the infinite branching endpoint families, namely those of D-type, fit in this picture.

In addition to the row invariant values detailed in Table~\ref{t:largeNLimits}, there are two other row invariant quantities: the residues of $f(n)$ and $1/f(n)$ at their poles. The latter precisely match the fraction determinant values as shown in Table~\ref{t:residues}. The correspondence is more than set-wise. Rather, we have a direct matching which reads
\begin{align}\label{eq:resId}
[\text{Res}_{n=n_c}(\frac{q}{p}(\alpha A_n \beta))]^{-1} = \sqrt{-\det(\frac{p}{q}(\overline{\beta} A_n \overline{\alpha}))} \ ,
\end{align}
with each side depending only on $\beta.$ This identity lets one easily check whether $\det(f(n))$ and $\det(\overline{f}(n))$ both are negated squares of the first six positive integers without first computing both $f(\overline{\gamma_n})$ and $f(\gamma_n).$ Considering the integer parameters yielding valid $f(n)$ allows us to determine those cases which have $\det (f(n))$ and $\det(\overline{f}(n))$ both meeting this determinant condition. Upon constraining to those cases obeying~\ref{eq:fracDetRelation} we find precisely those $f(n)$ in contact with 6D SCFT bases as we discuss shortly in greater detail.
{\renewcommand*{\arraystretch}{1.8}
     \begin{table}[htbp]
    \begin{center}
    {      \fontsize{8.2}{10.4}\selectfont\setlength{\tabcolsep}{8pt}
      \begin{tabular}{c|ccccccccccccc}
      $\overline{\alpha}:$ &  32222 & 3222 & 4 & 33 & 322 & 3 & 5 & 42 & 32 & 6 & 7 & $\emptyset$ \\
      \hline 
      $[\text{Res}_{n=n_c}(\frac{q}{p}(\alpha A_n \beta))]^{-1}$: & $6^2$ & $5^2$ & $3^2$ & $5^2$ & $4^2$ & $2^2$ & $4^2$ & $5^2$ & $3^2$ & $5^2$ & $6^2$ & $1$
     \end{tabular} }
        \caption{Inverses of residue value for $1/f(n)$ at its pole $n_c$ for $\alpha A_n \beta.$ Note the $\beta$ invariance.} 
        \label{t:residues}
           \end{center}
        \end{table}}

\noindent We now consider appropriate $n$ shifts so that
\begin{align}\label{eq:invertFrac}
f(\gamma_n) = \dfrac{k_u n + r_u}{k_l n + r_l}, \qquad \qquad  f(\overline{\gamma_{\widetilde{n}}}) = \dfrac{k_u n + r_u}{\widetilde{k_l} n + \widetilde{r_l}} \ 
\end{align}
holds. In most cases, this requires no modification from the minimally shifted expression. Rearranging~\eqref{eq:resId} yields
\begin{align}\label{eq:detsKuRelation}
\det(f(\gamma_{n})) \det(f(\overline{\gamma_{\widetilde{n}}})) = k_u^2 \ . 
\end{align}
Using~\eqref{eq:fracDetRelation} gives $\gcd(k_u,k_l) \cdot \gcd(k_u,\widetilde{k_l}) = k_u.$ 

This constraint allows for straightforward computation of $f(n)$ from $\overline{f}(n)$ and enables an alternate characterization of 6D SCFT base compatible $f(n).$ This also makes clear that unordered integer pairs $\{\det(f(n)),\text{Res}(f(n))\}$ are preserved by $T^*$ pairing (and endpoint reversal). We list these pairs explicitly as Table~\ref{t:intPairs} and find that $T^*$ pairs with matching unordered integer pairs appear to be precisely those with fixed $N_a$ size.

{
\renewcommand*{\arraystretch}{1.8}
     \begin{table}[htbp] 
    \begin{center}
    {      \fontsize{5.2}{6.8}\selectfont\setlength{\tabcolsep}{3pt}
      \begin{tabular}{c|cccccccccccc}
      \diaghead(1,-1){cccccccc}{$\ \alpha$}{$\beta \ $}& 7& 6& 5& 4& 42& 3& 33& 32& 322& 3222& 32222& 2 \\
      \hline 
$ 7 $ & $ \{6,6\} $ & $ \{5,6\} $ & $ \{4,6\} $ & $ \{3,6\} $ & $ \{5,6\} $ & $ \{2,6\} $ & $ \{5,6\} $ & $ \{3,6\} $ & $ \{4,6\} $ & $ \{5,6\} $ & $ \{6,6\} $ & $ \{1,6\} $ \\
 $
 6 $ & $ \{6,5\} $ & $ \{5,5\} $ & $ \{4,5\} $ & $ \{3,5\} $ & $ \{5,5\} $ & $ \{2,5\} $ & $ \{5,5\} $ & $ \{3,5\} $ & $ \{4,5\} $ & $ \{5,5\} $ & $ \{6,5\} $ & $ \{1,5\} $ \\
 $
 5 $ & $ \{6,4\} $ & $ \{5,4\} $ & $ \{4,4\} $ & $ \{3,4\} $ & $ \{5,4\} $ & $ \{2,4\} $ & $ \{5,4\} $ & $ \{3,4\} $ & $ \{4,4\} $ & $ \{5,4\} $ & $ \{6,4\} $ & $ \{1,4\} $ \\
 $
 4 $ & $ \{6,3\} $ & $ \{5,3\} $ & $ \{4,3\} $ & $ \{3,3\} $ & $ \{5,3\} $ & $ \{2,3\} $ & $ \{5,3\} $ & $ \{3,3\} $ & $ \{4,3\} $ & $ \{5,3\} $ & $ \{6,3\} $ & $ \{1,3\} $ \\
 $
 24 $ & $ \{6,5\} $ & $ \{5,5\} $ & $ \{4,5\} $ & $ \{3,5\} $ & $ \{5,5\} $ & $ \{2,5\} $ & $ \{5,5\} $ & $ \{3,5\} $ & $ \{4,5\} $ & $ \{5,5\} $ & $ \{6,5\} $ & $ \{1,5\} $ \\
 $
 3 $ & $ \{6,2\} $ & $ \{5,2\} $ & $ \{4,2\} $ & $ \{3,2\} $ & $ \{5,2\} $ & $ \{2,2\} $ & $ \{5,2\} $ & $ \{3,2\} $ & $ \{4,2\} $ & $ \{5,2\} $ & $ \{6,2\} $ & $ \{1,2\} $ \\
 $
 33 $ & $ \{6,5\} $ & $ \{5,5\} $ & $ \{4,5\} $ & $ \{3,5\} $ & $ \{5,5\} $ & $ \{2,5\} $ & $ \{5,5\} $ & $ \{3,5\} $ & $ \{4,5\} $ & $ \{5,5\} $ & $ \{6,5\} $ & $ \{1,5\} $ \\
 $
 23 $ & $ \{6,3\} $ & $ \{5,3\} $ & $ \{4,3\} $ & $ \{3,3\} $ & $ \{5,3\} $ & $ \{2,3\} $ & $ \{5,3\} $ & $ \{3,3\} $ & $ \{4,3\} $ & $ \{5,3\} $ & $ \{6,3\} $ & $ \{1,3\} $ \\
 $
 223 $ & $ \{6,4\} $ & $ \{5,4\} $ & $ \{4,4\} $ & $ \{3,4\} $ & $ \{5,4\} $ & $ \{2,4\} $ & $ \{5,4\} $ & $ \{3,4\} $ & $ \{4,4\} $ & $ \{5,4\} $ & $ \{6,4\} $ & $ \{1,4\} $ \\
 $
 2223 $ & $ \{6,5\} $ & $ \{5,5\} $ & $ \{4,5\} $ & $ \{3,5\} $ & $ \{5,5\} $ & $ \{2,5\} $ & $ \{5,5\} $ & $ \{3,5\} $ & $ \{4,5\} $ & $ \{5,5\} $ & $ \{6,5\} $ & $ \{1,5\} $ \\
 $
 22223 $ & $ \{6,6\} $ & $ \{5,6\} $ & $ \{4,6\} $ & $ \{3,6\} $ & $ \{5,6\} $ & $ \{2,6\} $ & $ \{5,6\} $ & $ \{3,6\} $ & $ \{4,6\} $ & $ \{5,6\} $ & $ \{6,6\} $ & $ \{1,6\} $ \\
 $
 2 $ & $ \{6,1\} $ & $ \{5,1\} $ & $ \{4,1\} $ & $ \{3,1\} $ & $ \{5,1\} $ & $ \{2,1\} $ & $ \{5,1\} $ & $ \{3,1\} $ & $ \{4,1\} $ & $ \{5,1\} $ & $ \{6,1\} $ & $ \{1,1\} $ \\
     \end{tabular} }
        \caption{Pairs $(\sqrt{-\det(f(n)}),\sqrt{-\text{Res}(1/f(n))}).$ Note that $T^*$ pairing is visible as reflection about the antidiagonal in the upper $11\times 11$ block and the $\alpha\sim 3$ entry in the final column preserves all unordered pairs.} 
        \label{t:intPairs}
           \end{center}
        \end{table}
 } 

\begin{subsection}{Relations determining SCFT compatible orbifold families}
In this section we discuss two characterizations of the rational functions $f(n)$ giving valid 6D SCFT endpoints. The first involves only the integers $1\leq k\leq6$ and relations on fraction determinants. The second involves the values which can arise in the limit $\lim_{n\to \infty} f(n).$ A key ingredient in our setup is that we only consider rational functions of the form~\eqref{eq:genFrac}.

In each case, precisely the minimally $n$ shifted endpoint family generators $f(n)$ are obtained by augmenting the constraint~\eqref{eq:genFrac}. Our aim is to step towards a parsimonious description of endpoint families short of requiring the Calabi-Yau condition on elliptic fibrations. Consider all $f(n)$ as in~\eqref{eq:genFrac} obeying~\eqref{eq:fracDetRelation} up to orientation of strings $m_1,\cdots, m_k$ (as in~\eqref{eq:contFrac}) and equivalence of the orbifold families they determine (i.e.\ by reducing the expressions $f(n)$ under consideration by minimally shifted $n$).\footnote{Note that certain orbifold families with $f(n)$ of the form~\eqref{eq:genFrac} are equivalent to others having distinct $f(n)$ by a shift of $n.$ We discard without loss of generality all $f(n)$ having non-minimal non-negative or negative values of $r_u,r_l.$} Next impose that $\det(f(n)) = -s^2$ for $1\leq s\leq 6$ an integer. This gives two simple sets of constraints which leave precisely the desired $f(n).$ 

Briefly, the following three requirements determine everything: 
\begin{itemize}
\item $\sqrt{-\det(f(n))}\in \{1,\cdots, 6\},$
\item $f(n)$ satisfy~\eqref{eq:fracDetRelation},
\item (a): $(\sqrt{\text{Res}(1/f(n))}=)\ \sqrt{-\det(\overline{f}(n))}\in \{1,\cdots, 6\}$ {\underline{or}} (b) Large $N$ limits must lie in $L_{\infty}.$
\end{itemize} 

\begin{subsubsection}{The permitted $f(n)$ via $\det(\overline{f}(n))$}
One approach is simply to require that $-\det(\overline{f}(n)), \ - \det(f(n))\in \{1^2, \cdots, 6^2\}.$ Note that it is convenient to compute the former as $[\text{Res}_{n=n_c}(\frac{1}{f(n)})]^{-1}.$ The trivial requirement that $q\leq p$ (as these determine equivalent orbifolds) in all remaining $f(n)$ can be taken without loss of generality. Explicit enumeration confirms that only precisely those $f(n)$ which correspond to 6D SCFT endpoint families remain.
\end{subsubsection}

\begin{subsubsection}{The permitted $f(n)$ via large $n$ limits}\label{s:linearEndpoints}
Alternatively, we can augment the determinant constraints with the minimal data set consisting of the 12 rationals giving permitted large $n$ limits. These $k_u/k_l$ values for $f(\alpha A_n \beta)$ are $\beta$ invariant (i.e.\ fixed on rows of Table~\ref{t:nFracs}, as are the aforementioned residues) and appear as Table~\ref{t:largeNLimits}. Let us refer to these twelve rationals as $L_{\infty}.$ Note that these are precisely the rationals $r=a/b \in \mathbb{Q}$ with $1\leq r,a,b \leq 6.$ As an aside, numerators of these row invariants match the corresponding column invariants $\sqrt{-\det(f(n))}$ appearing in Table~\ref{t:fracDet}. All $f(n)$ are fully determined by the values in $L_{\infty}$ with the requirement that $\lim_{n\to \infty} f(n) \in L_{\infty}.$ One can confirm this by explicit enumeration of the possible matches subject to the first two fraction determinant constraints above. This approach precisely recovers the permitted minimally shifted $f(n)$ that arise from 6D SCFT endpoints.

{\renewcommand*{\arraystretch}{1.8}
     \begin{table}[htbp]
    \begin{center}
    {      \fontsize{8.2}{10.4}\selectfont\setlength{\tabcolsep}{8pt}
      \begin{tabular}{c|ccccccccccccc}
      $\overline{\alpha}:$ &  32222 & 3222 & 4 & 33 & 322 & 3 & 5 & 42 & 32 & 6 & 7 & $\emptyset$ \\
      \hline 
      $\dfrac{k_u}{k_l}$: &  $\frac{6}{5}$&$\frac{5}{4}$&$\frac{3}{2}$&$\frac{5}{3}$&$4$&$2$&$\frac{4}{3}$&$\frac{5}{2}$&$3$&$5$&$6$&$1$
     \end{tabular} }
        \caption{Values of $\frac{k_u}{k_l}(\alpha)= \lim_{n\to\infty} f(\alpha A_n \beta).$} 
        \label{t:largeNLimits}
           \end{center}
        \end{table}
 }

\end{subsubsection}
\end{subsection}

\begin{subsection}{Intermediate determinants and canonical bases}\label{s:dets}
In~\cite{4DFtheory}, the intermediate quotient $\Gamma/H$ was discussed in relation to the truncations of bases for each endpoint string, where $H$ is the kernel of the map $\det: \Gamma \to U(1).$ In this section review this data and confirm it interacts sensibly with $T^*$ duality. We give the intermediate quotient $\Gamma/H$ for $H:=\ker(\det:\Gamma\to U(1))$ having order $m$ and $k/l$ giving the generator $e^{2 \pi i\cdot\frac{l}{k}}$ of $\Gamma/H.$ The values $(m,k/l)$ appeared in~\cite{4DFtheory}. Here we display them in terms of $n$ to give the minimally $n$ shifted $f(n)$ and lead ordering evidently respecting $T^*$ pairing.

We compute the values $m$ as $p/(\text{den} (\frac{q+1}{p}))$ where `$\text{den}$' denotes the denominator of a rational written in lowest terms. The values $l/k$ are then simply obtained as $l = \frac{q+1}{m}$ and $k = \frac{p}{m}.$ As noted in~\cite{4DFtheory} we have $\frac{p}{q+1}=m \frac{k}{l}$ with $k,l$ relatively prime. We observe that reflection about the antidiagonal in each block diagonal group (with $n$-coefficient of $m$ given by $d\in \{1,\cdots,6\}$) corresponds to endpoint reversal. Blocks consist precisely those endpoints built as $\alpha A_n\overline{\beta}$ from tails with matching fraction determinant, i.e.\ $\det(\alpha)=\det(\beta).$

{  \renewcommand*{\arraystretch}{1.8}
     \begin{table}[htbp] 
    \begin{center}
    {      \tiny \fontsize{4.1}{4.9}\selectfont \setlength{\tabcolsep}{1pt}
      \begin{tabular}{c|cccccccccccc}\diaghead(1,-1){cccccccccc}{$\alpha$}{$\ \beta $} & 32222 & 7 & 6 & 42 & 33 & 3222 & 5 & 322 & 4 & 32 & 3 & $\emptyset$ \\
      \hline 
$ 7 $ & $ (6 (n+1),6) $ & $ (2 (3 n+1),6) $ & $ (1,\frac{30 n+11}{5 n+2}) $ & $
   (1,\frac{30 n+17}{5 n+3}) $ & $ (1,\frac{30 n+23}{5 n+4}) $ & $
   (1,\frac{30 n+29}{5 n+5}) $ & $ (2,\frac{12 n+5}{2 n+1}) $ & $
   (2,\frac{12 n+11}{2 n+2}) $ & $ (1,\frac{18 n+9}{3 n+2}) $ & $
   (3,\frac{6 n+5}{n+1}) $ & $ (2,\frac{6 n+4}{n+1}) $ & $
   (1,\frac{6 n+1}{n+1}) $ \\
   $ 22223 $ & $ (2 (3 n+2),\frac{6}{5}) $ & $ (6 (n+1),\frac{6}{5}) $ & $
   (1,\frac{30 n+1}{25 n+1}) $ & $ (1,\frac{30 n+7}{25 n+6}) $ & $
   (1,\frac{30 n+13}{25 n+11}) $ & $ (1,\frac{30 n+19}{25 n+16}) $ & $
   (2,\frac{12 n+1}{10 n+1}) $ & $ (2,\frac{12 n+7}{10 n+6}) $ & $
   (3,\frac{6 n+1}{5 n+1}) $ & $ (1,\frac{18 n+9}{15 n+8}) $ & $
   (2,\frac{6 n+2}{5 n+2}) $ & $ (1,\frac{6 n+5}{5 n+5}) $ \\
  $
 2223 $ & $ (1,\frac{30 n+19}{24 n+15}) $ & $ (1,\frac{30 n+29}{24 n+23})
   $ & $ (5 (n+1),\frac{5}{4}) $ & $ (5 n+1,\frac{5}{4}) $ & $ (5
   n+2,\frac{5}{4}) $ & $ (5 n+3,\frac{5}{4}) $ & $ (1,\frac{20 n+1}{16
   n+1}) $ & $ (1,\frac{20 n+11}{16 n+9}) $ & $ (1,\frac{15 n+2}{12
   n+2}) $ & $ (1,\frac{15 n+7}{12 n+6}) $ & $ (1,\frac{10 n+3}{8
   n+3}) $ & $ (1,\frac{5 n+4}{4 n+4}) $ \\
  $
 33 $ & $ (1,\frac{30 n+13}{12 n+5}) $ & $ (1,\frac{30 n+23}{12 n+9}) $ & $
   (5 n+4,\frac{5}{2}) $ & $ (5 (n+1),\frac{5}{2}) $ & $ (5
   n+1,\frac{5}{2}) $ & $ (5 n+2,\frac{5}{2}) $ & $ (1,\frac{20 n+17}{8
   n+7}) $ & $ (1,\frac{20 n+7}{8 n+3}) $ & $ (1,\frac{15 n+14}{6
   n+6}) $ & $ (1,\frac{15 n+4}{6 n+2}) $ & $ (1,\frac{10 n+1}{4
   n+1}) $ & $ (1,\frac{5 n+3}{2 n+2}) $ \\
  $
 24 $ & $ (1,\frac{30 n+7}{18 n+4}) $ & $ (1,\frac{30 n+17}{18 n+10}) $ & $
   (5 n+3,\frac{5}{3}) $ & $ (5 n+4,\frac{5}{3}) $ & $ (5
   (n+1),\frac{5}{3}) $ & $ (5 n+1,\frac{5}{3}) $ & $ (1,\frac{20
   n+13}{12 n+8}) $ & $ (1,\frac{20 n+3}{12 n+2}) $ & $ (1,\frac{15
   n+11}{9 n+7}) $ & $ (1,\frac{15 n+1}{9 n+1}) $ & $ (1,\frac{10
   n+9}{6 n+6}) $ & $ (1,\frac{5 n+2}{3 n+2}) $ \\
  $
 6 $ & $ (1,\frac{30 n+31}{6 n+6}) $ & $ (1,\frac{30 n+11}{6 n+2}) $ & $
   (5 n+2,5) $ & $ (5 n+3,5) $ & $ (5 n+4,5) $ & $ (5 (n+1),5) $ & $ (1,\frac{20 n+9}{4
   n+2}) $ & $ (1,\frac{20 n+19}{4 n+4}) $ & $ (1,\frac{15 n+8}{3
   n+2}) $ & $ (1,\frac{15 n+13}{3 n+3}) $ & $ (1,\frac{10 n+7}{2
   n+2}) $ & $ (1,\frac{5 n+1}{n+1}) $ \\
  $
 223 $ & $ (2,\frac{12 n+7}{9 n+5}) $ & $ (2,\frac{12 n+11}{9 n+8}) $ & $
   (1,\frac{20 n+19}{15 n+14}) $ & $ (1,\frac{20 n+3}{15 n+2}) $ & $
   (1,\frac{20 n+7}{15 n+5}) $ & $ (1,\frac{20 n+11}{15 n+8}) $ & $
   (4 (n+1),\frac{4}{3}) $ & $ (2 (2 n+1),\frac{4}{3}) $ & $
   (1,\frac{12 n+1}{9 n+1}) $ & $ (1,\frac{12 n+5}{9 n+4}) $ & $
   (2,\frac{4 n+1}{3 n+1}) $ & $ (1,\frac{4 n+3}{3 n+3}) $ \\
  $
 5 $ & $ (2,\frac{12 n+13}{3 n+3}) $ & $ (2,\frac{12 n+5}{3 n+1}) $ & $
   (1,\frac{20 n+9}{5 n+2}) $ & $ (1,\frac{20 n+13}{5 n+3}) $ & $
   (1,\frac{20 n+17}{5 n+4}) $ & $ (1,\frac{20 n+21}{5 n+5}) $ & $ (2
   (2 n+1),4) $ & $ (4 (n+1),4) $ & $ (1,\frac{12 n+7}{3 n+2}) $ & $
   (1,\frac{12 n+11}{3 n+3}) $ & $ (2,\frac{4 n+3}{n+1}) $ & $
   (1,\frac{4 n+1}{n+1}) $ \\
  $
 23 $ & $ (1,\frac{18 n+9}{12 n+5}) $ & $ (3,\frac{6 n+5}{4 n+3}) $ & $
   (1,\frac{15 n+13}{10 n+8}) $ & $ (1,\frac{15 n+16}{10 n+10}) $ & $
   (1,\frac{15 n+4}{10 n+2}) $ & $ (1,\frac{15 n+7}{10 n+4}) $ & $
   (1,\frac{12 n+11}{8 n+7}) $ & $ (1,\frac{12 n+5}{8 n+3}) $ & $
   (3 (n+1),\frac{3}{2}) $ & $ (3 n+1,\frac{3}{2}) $ & $
   (1,\frac{6 n+1}{4 n+1}) $ & $ (1,\frac{3 n+2}{2 n+2}) $ \\
  $
 4 $ & $ (3,\frac{6 n+7}{2 n+2}) $ & $ (1,\frac{18 n+9}{6 n+2}) $ & $
   (1,\frac{15 n+8}{5 n+2}) $ & $ (1,\frac{15 n+11}{5 n+3}) $ & $
   (1,\frac{15 n+14}{5 n+4}) $ & $ (1,\frac{15 n+17}{5 n+5}) $ & $
   (1,\frac{12 n+7}{4 n+2}) $ & $ (1,\frac{12 n+13}{4 n+4}) $ & $ (3
   n+2,3) $ & $ (3 (n+1),3) $ & $ (1,\frac{6 n+5}{2 n+2}) $ & $ (1,\frac{3
   n+1}{n+1}) $ \\
  $
 3 $ & $ (2,\frac{6 n+8}{3 n+3}) $ & $ (2,\frac{6 n+4}{3 n+1}) $ & $
   (1,\frac{10 n+7}{5 n+2}) $ & $ (1,\frac{10 n+9}{5 n+3}) $ & $
   (1,\frac{10 n+11}{5 n+4}) $ & $ (1,\frac{10 n+13}{5 n+5}) $ & $
   (2,\frac{4 n+3}{2 n+1}) $ & $ (2,\frac{4 n+5}{2 n+2}) $ & $
   (1,\frac{6 n+5}{3 n+2}) $ & $ (1,\frac{6 n+7}{3 n+3}) $ & $ (2
   (n+1),2) $ & $ (1,\frac{2 n+1}{n+1}) $ \\
  $
 \emptyset $ & $ (1,\frac{6 n+11}{6 n+6}) $ & $ (1,\frac{6 n+7}{6 n+2}) $ & $
   (1,\frac{5 n+6}{5 n+2}) $ & $ (1,\frac{5 n+7}{5 n+3}) $ & $
   (1,\frac{5 n+8}{5 n+4}) $ & $ (1,\frac{5 n+9}{5 n+5}) $ & $
   (1,\frac{4 n+5}{4 n+2}) $ & $ (1,\frac{4 n+7}{4 n+4}) $ & $
   (1,\frac{3 n+4}{3 n+2}) $ & $ (1,\frac{3 n+5}{3 n+3}) $ & $
   (1,\frac{2 n+3}{2 n+2}) $ & $ (n+1,1) $ \\
        \end{tabular} }
        \caption{Values of $(m,k/l)$ giving the order of $H:=\ker(\det:\Gamma\to U(1))$ and generator $e^{2 \pi i\cdot\frac{l}{k}}$ of the intermediate group $\Gamma/H$. Block groupings respect the $T^*$ pairing  appearing here as transposition of the grid.} 
        \label{t:intDets}
           \end{center}
        \end{table}
 }

Minimal resolutions of each endpoint give one distinguished class of bases as discussed in~\cite{atomic} by blowing up as little as possible to obtain a valid F-theory base in each endpoint. One sided truncations of the possible minimal resolutions are known to comprise the available block diagonal endpoint groupings of the intermediate determinant map $\det:\Gamma \to U(1)$ as discussed in~\cite{4DFtheory} and summarized in Table~\ref{t:diagTruncs}. However, the remaining endpoints require moving slightly beyond minimal resolution to give a similar characterization. (Consolidating instantons after blowing up slightly past a minimal resolution makes the relationship between common determinant behavior groups and truncation behavior of resolutions evident as a generalization of Table~\ref{t:diagTruncs}.)
{\renewcommand*{\arraystretch}{1.8}
\begin{table}
\begin{center}
{\tiny \fontsize{5.1}{6.1}\selectfont \setlength{\tabcolsep}{3pt}
      \begin{tabular}{c|cccccccccccc}\diaghead(1,-1){cccccccccc}{$d$}{$\ \gamma $} \\
      \hline 
6 & $2231513221\overset{n+1}{(\overbrace{\{12\}12231513221)\cdots}}\{12\}1223151322 $\\
5 & $231513221\overset{n+1}{(\overbrace{\{12\}12231513221)\cdots}}\{12\}122315132$ \\
4 & $2321\overset{n+1}{\overbrace{(812321)\cdots}}81232$\\
3 & $31\overset{n+1}{\overbrace{(6131)\cdots}}613$ \\
2 & $\overset{n+1}{\overbrace{(41)\cdots}}4$ \\
1 & $\overset{n+1}{\overbrace{(2)\cdots}}\emptyset$
            \end{tabular} }
        \caption{Level $n$ bases $\gamma$ yielding block diagonal endpoint groups with common $\det:\Gamma \to U(1)$ behavior with $n$ coefficient of $m$ given by $d.$ The terms in parentheses are repeated $n+1$ times.} 
        \label{t:diagTruncs}
        \end{center}
        \end{table}}

The collection of near minimal resolutions is structurally somewhat unwieldy in comparison with a certain class of linear endpoint base representatives we will term {\it canonical endpoint representatives} or {\it canonical bases} obtained as the truncations of a single (infinite) chain of curves. 
As recently reported in~\cite{heckman2019top}, these biject with linear endpoints to cleanly characterize the $12\times12$ endpoint family structure as left and right truncations of the bases
\begin{align}\label{eq:theBase} 
\overset{L_l}{\overbrace{12231513221}}\ \overset{n+1}{\overbrace{(\langle 12 \rangle 12231513221) \cdots}}\langle 12\rangle \ \overset{L_r}{\overbrace{12231513221}} \ .
\end{align}
This structure induces a unique canonical ordering of leads/tails respecting $T^*$ pairing and common intermediate determinant behavior in each block diagonal group as is evident in Table~\ref{t:intDets}. Rows and columns correspond to one sided truncations of a fixed canonical base, transposition to endpoint reversal, and $T^*$ pairing to reflection about the $11\times 11$ block antidiagonal. 

Inclusion induced structure endows each level-$n$ endpoint set with a unique canonical partial order compatible with the truncation induced lead ordering and $T^*$ compatible symmetric lead ordering. The left/right truncations of $L_l,L_r$ in the fixed-$n$ base in~\eqref{eq:theBase} yields the level $n$ endpoints by blowing down with an $n$-shift obtained by replacing the empty lead by a single $-2$ curve. This places a natural level relation between endpoint families (i.e.\ a natural family-wise fixed choice of shift $n\to \widetilde{n}$ for each). The inclusion structure places a partial ordering on endpoints compatible with the cyclic ordering on leads yielding our choice in~\eqref{eq:leads}
arising via increasingly truncated strings in~\eqref{eq:theBase}.
This truncation induced lead order is compatible with $T^*$ duality. As above, $T^*$ pairs of endpoint families appear as as the usual reflection in this partial order. Roots of $p(n)$ modulo 1 form uniquely large groups of shared signs respecting the $T^*$ induced sign swap, as we saw in Table~\ref{t:reducedPRootsTruncOrder}. Further relationships hold between $T^*$ paired endpoints and the structure of bases $b_{l,r,n}$ with form~\eqref{eq:theBase}. The $T^*$ paired bases $b_{i,j,n},b_{i,j,n}^*$ have the following two properties.
\begin{itemize}
\item The combined total number of curves in each pair is constant for fixed $n.$
\item The paired bases permit a gluing by introducing a single curve.
\end{itemize}
Each follows from the relationship of $T^*$ pairing with endpoint/canonical base pairing. Together these make clear that $T^*$ pairing respects a simple truncation structure with each $T^*$ pair naturally associated to a unique longer base and hence (distinct) endpoint.  

Regularity in the structure of linear 6D SCFT endpoint families is revealed by the above truncation structure. A graph can be defined on endpoint families by joining those that contain bases of the form~\eqref{eq:theBase} which are minimally truncation related, i.e.\ differing by a single curve. This graph $\Gamma_e$ is illustrated in Figure~\ref{f:endpointFamilies}. It has automorphism group $G\cong D_{12}$ of order 24 with orbits of sizes $12^6,6$ (and generators acting by (i) separately cycling the 6 central points and the radial bands of shared lead simultaneously and (ii,iii) reflection about an opposite pair of outermost vertices where for (ii) these are 4-valent and leave 8 fixed points while for (iii) these are 2-valent and leave 6 fixed points). The order two generator $g_a$ of type (ii) above corresponds to $T^*$ pairing (this being the reflection about the line from $v_{1,a}\sim3A_n3$ to $v_{1,a}\sim \emptyset A_n\emptyset$). The type (iii) generator $g_b$ arises as the reflection about the line between $v_{1,b}\sim 223A_n3222$ and $v_{2,b}\sim5A_n4.$ Note that a less symmetric graph having automorphism group of order 2 results unless we consider multiple levels in this construction.

Forms of $f(n)$ and $\det:\Gamma\to U(1)$ invariants with this induced lead ordering appear in Table,~\ref{t:nFracs},\ref{t:truncDets}, respectively. Note that the $T^*$ pairing respects the $n$ coefficients appearing in $m,k.$

{  \renewcommand*{\arraystretch}{1.8}
     \begin{table}[htbp] 
    \begin{center}
    {      \tiny \fontsize{4.1}{4.9}\selectfont \setlength{\tabcolsep}{1pt}
      \begin{tabular}{c|cccccccccccc}\diaghead(1,-1){cccccccccc}{$\alpha$}{$\ \beta $}&
  32222 & 3222 & 322 & 32 & 33 & 3 & 42 & 4 & 5 & 6 & 7 & $\emptyset$\\
      \hline 
 $22223 $ & $ (2 (3 n+2),\frac{6}{5})$ & $ (1,\frac{30 n+19}{25 n+16})$ & $ (2,\frac{12 n+7}{10 n+6})$ & $ (1,\frac{18 n+9}{15 n+8})$ & $ (1,\frac{30 n+13}{25 n+11})$ & $ (2,\frac{6 n+2}{5 n+2})$ & $ (1,\frac{30 n+7}{25 n+6})$ & $ (3,\frac{6 n+1}{5 n+1})$ & $ (2,\frac{12 n+1}{10 n+1})$ & $
   (1,\frac{30 n+1}{25 n+1})$ & $ (6 (n+1),\frac{6}{5})$ & $ (1,\frac{6 n+5}{5 n+5})$ \\ $
 2223 $ & $ (1,\frac{30 n+19}{24 n+15})$ & $ (5 n+3,\frac{5}{4})$ & $ (1,\frac{20 n+11}{16 n+9})$ & $ (1,\frac{15 n+7}{12 n+6})$ & $ (5 n+2,\frac{5}{4})$ & $ (1,\frac{10 n+3}{8 n+3})$ & $ (5 n+1,\frac{5}{4})$ & $ (1,\frac{15 n+2}{12 n+2})$ & $ (1,\frac{20 n+1}{16 n+1})$ & $ (5
   (n+1),\frac{5}{4})$ & $ (1,\frac{30 n+29}{24 n+23})$ & $ (1,\frac{5 n+4}{4 n+4})$ \\ $
 223 $ & $ (2,\frac{12 n+7}{9 n+5})$ & $ (1,\frac{20 n+11}{15 n+8})$ & $ (2 (2 n+1),\frac{4}{3})$ & $ (1,\frac{12 n+5}{9 n+4})$ & $ (1,\frac{20 n+7}{15 n+5})$ & $ (2,\frac{4 n+1}{3 n+1})$ & $ (1,\frac{20 n+3}{15 n+2})$ & $ (1,\frac{12 n+1}{9 n+1})$ & $ (4 (n+1),\frac{4}{3})$ & $
   (1,\frac{20 n+19}{15 n+14})$ & $ (2,\frac{12 n+11}{9 n+8})$ & $ (1,\frac{4 n+3}{3 n+3})$ \\ $
 23 $ & $ (1,\frac{18 n+9}{12 n+5})$ & $ (1,\frac{15 n+7}{10 n+4})$ & $ (1,\frac{12 n+5}{8 n+3})$ & $ (3 n+1,\frac{3}{2})$ & $ (1,\frac{15 n+4}{10 n+2})$ & $ (1,\frac{6 n+1}{4 n+1})$ & $ (1,\frac{15 n+16}{10 n+10})$ & $ (3 (n+1),\frac{3}{2})$ & $ (1,\frac{12 n+11}{8 n+7})$ & $
   (1,\frac{15 n+13}{10 n+8})$ & $ (3,\frac{6 n+5}{4 n+3})$ & $ (1,\frac{3 n+2}{2 n+2})$ \\ $
 33 $ & $ (1,\frac{30 n+13}{12 n+5})$ & $ (5 n+2,\frac{5}{2})$ & $ (1,\frac{20 n+7}{8 n+3})$ & $ (1,\frac{15 n+4}{6 n+2})$ & $ (5 n+1,\frac{5}{2})$ & $ (1,\frac{10 n+1}{4 n+1})$ & $ (5 (n+1),\frac{5}{2})$ & $ (1,\frac{15 n+14}{6 n+6})$ & $ (1,\frac{20 n+17}{8 n+7})$ & $ (5
   n+4,\frac{5}{2})$ & $ (1,\frac{30 n+23}{12 n+9})$ & $ (1,\frac{5 n+3}{2 n+2})$ \\ $
 3 $ & $ (2,\frac{6 n+8}{3 n+3})$ & $ (1,\frac{10 n+13}{5 n+5})$ & $ (2,\frac{4 n+5}{2 n+2})$ & $ (1,\frac{6 n+7}{3 n+3})$ & $ (1,\frac{10 n+11}{5 n+4})$ & $ (2 (n+1),2) $ & $ (1,\frac{10 n+9}{5 n+3})$ & $ (1,\frac{6 n+5}{3 n+2})$ & $ (2,\frac{4 n+3}{2 n+1})$ & $ (1,\frac{10 n+7}{5 n+2})$ & $
   (2,\frac{6 n+4}{3 n+1})$ & $ (1,\frac{2 n+1}{n+1})$ \\ $
 24 $ & $ (1,\frac{30 n+7}{18 n+4})$ & $ (5 n+1,\frac{5}{3})$ & $ (1,\frac{20 n+3}{12 n+2})$ & $ (1,\frac{15 n+1}{9 n+1})$ & $ (5 (n+1),\frac{5}{3})$ & $ (1,\frac{10 n+9}{6 n+6})$ & $ (5 n+4,\frac{5}{3})$ & $ (1,\frac{15 n+11}{9 n+7})$ & $ (1,\frac{20 n+13}{12 n+8})$ & $ (5
   n+3,\frac{5}{3})$ & $ (1,\frac{30 n+17}{18 n+10})$ & $ (1,\frac{5 n+2}{3 n+2})$ \\ $
 4 $ & $ (3,\frac{6 n+7}{2 n+2})$ & $ (1,\frac{15 n+17}{5 n+5})$ & $ (1,\frac{12 n+13}{4 n+4})$ & $ (3 (n+1),3) $ & $ (1,\frac{15 n+14}{5 n+4})$ & $ (1,\frac{6 n+5}{2 n+2})$ & $ (1,\frac{15 n+11}{5 n+3})$ & $ (3 n+2,3) $ & $ (1,\frac{12 n+7}{4 n+2})$ & $ (1,\frac{15 n+8}{5 n+2})$ & $ (1,\frac{18
   n+9}{6 n+2})$ & $ (1,\frac{3 n+1}{n+1})$ \\ $
 5 $ & $ (2,\frac{12 n+13}{3 n+3})$ & $ (1,\frac{20 n+21}{5 n+5})$ & $ (4 (n+1),4) $ & $ (1,\frac{12 n+11}{3 n+3})$ & $ (1,\frac{20 n+17}{5 n+4})$ & $ (2,\frac{4 n+3}{n+1})$ & $ (1,\frac{20 n+13}{5 n+3})$ & $ (1,\frac{12 n+7}{3 n+2})$ & $ (2 (2 n+1),4) $ & $ (1,\frac{20 n+9}{5 n+2})$ & $ (2,\frac{12
   n+5}{3 n+1})$ & $ (1,\frac{4 n+1}{n+1})$ \\ $
 6 $ & $ (1,\frac{30 n+31}{6 n+6})$ & $ (5 (n+1),5) $ & $ (1,\frac{20 n+19}{4 n+4})$ & $ (1,\frac{15 n+13}{3 n+3})$ & $ (5 n+4,5) $ & $ (1,\frac{10 n+7}{2 n+2})$ & $ (5 n+3,5) $ & $ (1,\frac{15 n+8}{3 n+2})$ & $ (1,\frac{20 n+9}{4 n+2})$ & $ (5 n+2,5) $ & $ (1,\frac{30 n+11}{6 n+2})$ & $ (1,\frac{5
   n+1}{n+1})$ \\ $
 7 $ & $ (6 (n+1),6) $ & $ (1,\frac{30 n+29}{5 n+5})$ & $ (2,\frac{12 n+11}{2 n+2})$ & $ (3,\frac{6 n+5}{n+1})$ & $ (1,\frac{30 n+23}{5 n+4})$ & $ (2,\frac{6 n+4}{n+1})$ & $ (1,\frac{30 n+17}{5 n+3})$ & $ (1,\frac{18 n+9}{3 n+2})$ & $ (2,\frac{12 n+5}{2 n+1})$ & $ (1,\frac{30 n+11}{5 n+2})
   $ & $ (2 (3 n+1),6) $ & $ (1,\frac{6 n+1}{n+1})$ \\ $
 \emptyset $ & $ (1,\frac{6 n+11}{6 n+6})$ & $ (1,\frac{5 n+9}{5 n+5})$ & $ (1,\frac{4 n+7}{4 n+4})$ & $ (1,\frac{3 n+5}{3 n+3})$ & $ (1,\frac{5 n+8}{5 n+4})$ & $ (1,\frac{2 n+3}{2 n+2})$ & $ (1,\frac{5 n+7}{5 n+3})$ & $ (1,\frac{3 n+4}{3 n+2})$ & $ (1,\frac{4 n+5}{4 n+2})$ & $ (1,\frac{5
   n+6}{5 n+2})$ & $ (1,\frac{6 n+7}{6 n+2})$ & $ (n+1,1)$\\
      \end{tabular} }
        \caption{Truncation induced ordering of the values of $(m,k/l)$ describing $\det:\Gamma\to U(1).$} 
        \label{t:truncDets}
           \end{center}
        \end{table}
 }

\begin{figure}
\begin{center}
\includegraphics[width=13cm]{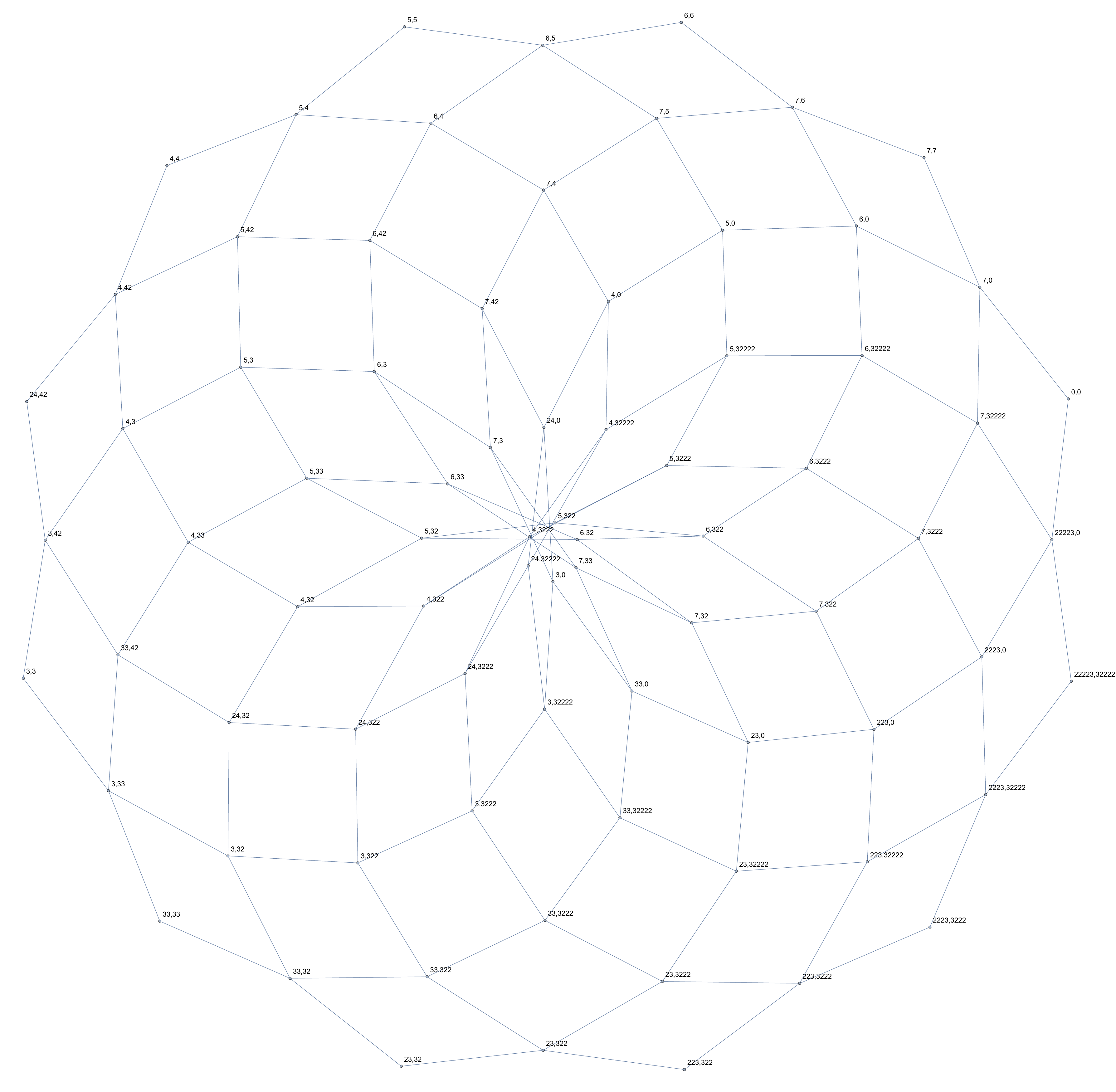}
\end{center}
\caption{Linear 6D SCFT endpoint families with edges joining nearest truncation of canonical base representatives with form~\eqref{eq:theBase}.}
\label{f:endpointFamilies}
\end{figure}

We can similarly define truncation induced graphs from the subcollection of endpoints that blow down from truncations of $L_l,L_r$ in
\begin{align}\label{eq:theBase2} 
b_{l,r,n}\sim\overset{L_l}{\overbrace{12321}}\ \overset{n}{\overbrace{(\langle 8 \rangle 12321) \cdots}}\langle 8\rangle \ \overset{L_r}{\overbrace{12321}} \ . 
\end{align}
This analogously yields a graph with automorphism group $G\cong D_6$ again with one generator of order two corresponding to $T^*$ pairing. 
\end{subsection}

\end{section}

\begin{section}{Conclusions}

The landscape of 6D SCFT F-theory models has a rich structure that has allowed tools from physics to come to bear on a variety of combinatorial problems. Here, we have studied macro-scale features of this landscape with a focus on families of theories with a shared endpoint in hopes to fuel this progress. The novel SCFT family duality we have outlined is of particular interest since it arises naturally as the only possible level inversion of the endpoint tower respecting partial ordering of endpoints induced by 6D RG flow. The matching we find between the established nilpotent orbit groupings of endpoints and the novel level invariant algebraic quantities of rational functions $f(n)$ that endpoint families illustrates the significant degree to which purely algebraic properties of the functions characterizing $\C^2/\Gamma$ orbifold families control structure of 6D SCFT F-theory models over these bases. We have also seen that these functions can be given a comparatively clean characterization. We hope that future work will enable an extension of our analysis to better understand the combinatorics of this collection of bases and link it with known mathematical structures including CY threefold moduli space and finite groups acting naturally on endpoint collections. Of particular interest along these lines is the $\Z^2$ lattice extension of the truncation induced partial order on bases that has appeared here via extrapolation of endpoint families to negative levels. 

We have confined our discussion here primarily to linear bases and endpoints for computational ease. It would be instructive to determine to what degree perfect matchings of SCFT families are achieved by removing this limitation. Nonetheless, we have seen that such matchings typically do exist and are induced via addition of a uniform gauge summand to the collections SCFTs in endpoints paired by $T^*$ duality. Still, a systematic analysis of these summands remains to be carried out. We hope that the pairing detailed here and its role as inversion of level in the tower of 6D SCFTs may help through study of $T^*$ paired gauge algebra structure to provide checks on the classification of 6D SCFTs, aid in classifying 6D RG flows, and inform attempts to characterize the landscape of SCFTs in fewer than six dimensions.

\end{section}

\acknowledgments We would like to thank David R.\ Morrison, Marco Bertolini, Tom Rudelius, and Alessandro Tomasiello for valuable discussions. We are grateful to the University of California, Santa Barbara Department of Mathematics, the Department of Mathematics and Computer Science at Wesleyan University, and the Department of Mathematics at the University of California, Davis for hospitality during the course of this work.

\appendix 

\begin{section}{Extrapolated endpoints for $n<0$}\label{s:extraps}
In this section we detail the extrapolation of endpoints via their defining rational functions to infinitely many negative values of the level parameter. For such values, the same skyscraper of endpoints we have seen for $n>0$ reemerges with a $T^*$ reflection within each family. A novel combinatorial structure appears in the gap for small $n$ before this $T^*$ dual behavior initiates. Infinitely many extrapolated endpoints appear, but some care is required to arrive at this conclusion. Crucially, we do not merely obtain rational numbers via the functions $f(n)$ of Table~\ref{t:nFracs} but instead carefully extrapolate a consistent rational by tracking numerator and denominator signs obtained from $f(n)$ for a smooth extrapolation matching the underlying orbifold action in~\eqref{eq:orbAction}.

Recall that all linear 6D SCFT endpoints appear in association with rational numbers obtained from the rational functions appearing in Table~\ref{t:nFracs} for some $n\geq -1$ as shown in~\cite{4DFtheory}. The association given via~\eqref{eq:contFrac} for linear endpoints becomes slightly more complicated for branching endpoints, as reviewed in Section~\ref{s:dTypes}. For this reason, we confine our discussion here to linear endpoints.

To carry out extrapolations that respect endpoint reversal and persist to arbitrarily negative level, we follow~\eqref{eq:orbAction} in place of aiming for a positive rational to appear directly for use in~\eqref{eq:contFrac}. The twelve $T^*$ self-dual families has a single level in which $p(n)$ becomes zero. This appears to prevent any meaningful extrapolation. In all remaining cases, the following method provides extrapolated endpoints that we collect in Table~\ref{t:extraps}. 

{ 
     \begin{table}[htbp] 
     \centering
    \begin{center}
    \centering
    \begin{subtable}{.9\linewidth}
    {      \tiny \setlength{\tabcolsep}{3pt}
      \begin{tabular}{c|cccccccccccc}\diaghead(1,-1){cccccccc}{$\ \alpha$}{$\beta \ $} & 32222 & 3222 & 322 & 32 & 33 & 3 & 42 & 4 & 5 & 6 & 7 & $\emptyset$ \\
            \hline 
 22223 & 222242222 & 22224222 & 2222422 & 222242 & 222243 & 22224 & 222252 & 22225 & 22226 & 22227 & 22228 & 2222 \\
 2223 & 22242222 & 2224222 & 222422 & 22242 & 22243 & 2224 & 22252 & 2225 & 2226 & 2227 & 2228 & 222 \\
 223 & 2242222 & 224222 & 22422 & 2242 & 2243 & 224 & 2252 & 225 & 226 & 227 & 228 & 22 \\
 23 & 242222 & 24222 & 2422 & 242 & 243 & 24 & 252 & 25 & 26 & 27 & 28 & 2 \\
 33 & 342222 & 34222 & 3422 & 342 & 343 & 34 & 352 & 35 & 36 & 37 & 38 & 3 \\
 3 & 42222 & 4222 & 422 & 42 & 43 & 4 & 52 & 5 & 6 & 7 & 8 & $\emptyset$ \\
 24 & 252222 & 25222 & 2522 & 252 & 253 & 25 & 262 & 26 & 27 & 28 & 29 & 2 \\
 4 & 52222 & 5222 & 522 & 52 & 53 & 5 & 62 & 6 & 7 & 8 & 9 & $\emptyset$ \\
 5 & 62222 & 6222 & 622 & 62 & 63 & 6 & 72 & 7 & 8 & 9 & (10) & $\emptyset$ \\
 6 & 72222 & 7222 & 722 & 72 & 73 & 7 & 82 & 8 & 9 & (10) & (11) & $\emptyset$ \\
 7 & 82222 & 8222 & 822 & 82 & 83 & 8 & 92 & 9 & (10) & (11) & (12) & $\emptyset$ \\
$\emptyset$ & 2222 & 222 & 22 & 2 & 3 & $\emptyset$ & 2 & $\emptyset$ & $\emptyset$ & $\emptyset$ & $\emptyset$ & - \\
     \end{tabular} }
        \caption{Level $n=-1$.} 
        \label{t:extrapMinus1}
        \end{subtable}
           \end{center}
\centering           
         \begin{subtable}{1\linewidth}
         \centering
    \begin{center}
    {      \tiny \setlength{\tabcolsep}{3pt}
      \begin{tabular}{c|cccccccccccc}\diaghead(1,-1){cccccccc}{$\ \alpha$}{$\beta \ $} & 32222 & 3222 & 322 & 32 & 33 & 3 & 42 & 4 & 5 & 6 & 7 & $\emptyset$ \\
            \hline 
 22223 & 2223222 & 222322 & 22232 & 2223 & 2224 & 222 & 223 & 22 & 2 & $\emptyset$ & - & $\emptyset$ \\
 2223 & 223222 & 22322 & 2232 & 223 & 224 & 22 & 23 & 2 & $\emptyset$ & - & $\emptyset$ & $\emptyset$ \\
 223 & 23222 & 2322 & 232 & 23 & 24 & 2 & 3 & $\emptyset$ & - & $\emptyset$ & 2 & $\emptyset$ \\
 23 & 3222 & 322 & 32 & 3 & 4 & $\emptyset$ & $\emptyset$ & - & $\emptyset$ & 2 & 22 & $\emptyset$ \\
 33 & 4222 & 422 & 42 & 4 & 5 & $\emptyset$ & - & $\emptyset$ & 3 & 23 & 223 & 2 \\
 3 & 222 & 22 & 2 & $\emptyset$ & $\emptyset$ & - & $\emptyset$ & $\emptyset$ & 2 & 22 & 222 & $\emptyset$ \\
 24 & 322 & 32 & 3 & $\emptyset$ & - & $\emptyset$ & 5 & 4 & 24 & 224 & 2224 & 3 \\
 4 & 22 & 2 & $\emptyset$ & - & $\emptyset$ & $\emptyset$ & 4 & 3 & 23 & 223 & 2223 & 2 \\
 5 & 2 & $\emptyset$ & - & $\emptyset$ & 3 & 2 & 42 & 32 & 232 & 2232 & 22232 & 22 \\
 6 & $\emptyset$ & - & $\emptyset$ & 2 & 32 & 22 & 422 & 322 & 2322 & 22322 & 222322 & 222 \\
 7 & - & $\emptyset$ & 2 & 22 & 322 & 222 & 4222 & 3222 & 23222 & 223222 & 2223222 & 2222 \\
$\emptyset$  & $\emptyset$ & $\emptyset$ & $\emptyset$ & $\emptyset$ & 2 & $\emptyset$ & 3 & 2 & 22 & 222 & 2222 & $\emptyset$ \\
     \end{tabular} }
        \caption{Level $n=-2$.} 
        \label{t:extrapMinus2}
           \end{center}
           \end{subtable}
\centering
         \begin{subtable}{1\linewidth}         
    \begin{center}
    {      \tiny \setlength{\tabcolsep}{3pt}
      \begin{tabular}{c|cccccccccccc}\diaghead(1,-1){cccccccc}{$\ \alpha$}{$\beta \ $} & 32222 & 3222 & 322 & 32 & 33 & 3 & 42 & 4 & 5 & 6 & 7 & $\emptyset$ \\
            \hline 
 22223 & 77 & 67 & 57 & 47 & 247 & 37 & 337 & 237 & 2237 & 22237 & 222237 & 27 \\
 2223 & 76 & 66 & 56 & 46 & 246 & 36 & 336 & 236 & 2236 & 22236 & 222236 & 26 \\
 223 & 75 & 65 & 55 & 45 & 245 & 35 & 335 & 235 & 2235 & 22235 & 222235 & 25 \\
 23 & 74 & 64 & 54 & 44 & 244 & 34 & 334 & 234 & 2234 & 22234 & 222234 & 24 \\
 33 & 742 & 642 & 542 & 442 & 2442 & 342 & 3342 & 2342 & 22342 & 222342 & 2222342 & 242 \\
 3 & 73 & 63 & 53 & 43 & 243 & 33 & 333 & 233 & 2233 & 22233 & 222233 & 23 \\
 24 & 733 & 633 & 533 & 433 & 2433 & 333 & 3333 & 2333 & 22333 & 222333 & 2222333 & 233 \\
 4 & 732 & 632 & 532 & 432 & 2432 & 332 & 3332 & 2332 & 22332 & 222332 & 2222332 & 232 \\
 5 & 7322 & 6322 & 5322 & 4322 & 24322 & 3322 & 33322 & 23322 & 223322 & 2223322 & 22223322 & 2322 \\
 6 & 73222 & 63222 & 53222 & 43222 & 243222 & 33222 & 333222 & 233222 & 2233222 & 22233222 & 222233222 & 23222 \\
 7 & 732222 & 632222 & 532222 & 432222 & 2432222 & 332222 & 3332222 & 2332222 & 22332222 & 222332222 & 2222332222 & 232222 \\
 $\emptyset$ & 72 & 62 & 52 & 42 & 242 & 32 & 332 & 232 & 2232 & 22232 & 222232 & 22 \\
     \end{tabular} }
        \caption{Level $n=-3$.} 
        \label{t:extrapMinus3}
           \end{center}
           \end{subtable}
\centering
    \begin{subtable}{1.1\linewidth}
         \centering
    \begin{center}
    {      \tiny \setlength{\tabcolsep}{1pt}
           \hspace{-0.5in} \begin{tabular}{c|cccccccccccc}\diaghead(1,-1){cccccccc}{$\ \alpha$}{$\beta \ $} & 32222 & 3222 & 322 & 32 & 33 & 3 & 42 & 4 & 5 & 6 & 7 & $\emptyset$ \\
            \hline 
 
  22223 & 7$A_i$7 & 7$A_i$6 & 7$A_i$5 & 7$A_i$4 & 7$A_i$24 & 7$A_i$3 & 7$A_i$33 & 7$A_i$23 & 7$A_i$223 & 7$A_i$2223 & 7$A_i$22223 & 7$A_{i+1}$ \\
  2223 & 6$A_i$7 & 6$A_i$6 & 6$A_i$5 & 6$A_i$4 & 6$A_i$24 & 6$A_i$3 & 6$A_i$33 & 6$A_i$23 & 6$A_i$223 & 6$A_i$2223 & 6$A_i$22223 & 6$A_{i+1}$ \\
  223 & 5$A_i$7 & 5$A_i$6 & 5$A_i$5 & 5$A_i$4 & 5$A_i$24 & 5$A_i$3 & 5$A_i$33 & 5$A_i$23 & 5$A_i$223 & 5$A_i$2223 & 5$A_i$22223 & 5$A_{i+1}$ \\
  23 & 4$A_i$7 & 4$A_i$6 & 4$A_i$5 & 4$A_i$4 & 4$A_i$24 & 4$A_i$3 & 4$A_i$33 & 4$A_i$23 & 4$A_i$223 & 4$A_i$2223 & 4$A_i$22223 & 4$A_{i+1}$ \\
  33 & 24$A_i$7 & 24$A_i$6 & 24$A_i$5 & 24$A_i$4 & 24$A_i$24 & 24$A_i$3 & 24$A_i$33 & 24$A_i$23 & 24$A_i$223 & 24$A_i$2223 & 24$A_i$22223 & 24$A_{i+1}$ \\
  3 & 3$A_i$7 & 3$A_i$6 & 3$A_i$5 & 3$A_i$4 & 3$A_i$24 & 3$A_i$3 & 3$A_i$33 & 3$A_i$23 & 3$A_i$223 & 3$A_i$2223 & 3$A_i$22223 & 3$A_{i+1}$ \\
  24 & 33$A_i$7 & 33$A_i$6 & 33$A_i$5 & 33$A_i$4 & 33$A_i$24 & 33$A_i$3 & 33$A_i$33 & 33$A_i$23 & 33$A_i$223 & 33$A_i$2223 & 33$A_i$22223 & 33$A_{i+1}$ \\
  4 & 23$A_i$7 & 23$A_i$6 & 23$A_i$5 & 23$A_i$4 & 23$A_i$24 & 23$A_i$3 & 23$A_i$33 & 23$A_i$23 & 23$A_i$223 & 23$A_i$2223 & 23$A_i$22223 & 23$A_{i+1}$ \\
  5 & 223$A_i$7 & 223$A_i$6 & 223$A_i$5 & 223$A_i$4 & 223$A_i$24 & 223$A_i$3 & 223$A_i$33 & 223$A_i$23 & 223$A_i$223 & 223$A_i$2223 & 223$A_i$22223 & 223$A_{i+1}$ \\
  6 & 2223$A_i$7 & 2223$A_i$6 & 2223$A_i$5 & 2223$A_i$4 & 2223$A_i$24 & 2223$A_i$3 & 2223$A_i$33 & 2223$A_i$23 & 2223$A_i$223 & 2223$A_i$2223 & 2223$A_i$22223 & 2223$A_{i+1}$ \\
  7 & 22223$A_i$7 & 22223$A_i$6 & 22223$A_i$5 & 22223$A_i$4 & 22223$A_i$24 & 22223$A_i$3 & 22223$A_i$33 & 22223$A_i$23 & 22223$A_i$223 & 22223$A_i$2223 & 22223$A_i$22223 & 22223$A_{i+1}$ \\
 $\emptyset$  & $A_{i+1}$7 & $A_{i+1}$6 & $A_{i+1}$5 & $A_{i+1}$4 & $A_{i+1}$24 & $A_{i+1}$3 & $A_{i+1}$33 & $A_{i+1}$23 & $A_{i+1}$223 & $A_{i+1}$2223 & $A_{i+1}$22223 & $A_{i+1}$ \\
     \end{tabular}\caption{General level $n\leq-3$ where $i:=-(n+3).$ } 
     \label{t:extrapMinusGen} }
           \end{center}
        \end{subtable}  
        \caption{Endpoints extrapolated to $n<0$ levels via rational functions $f(n)$.}\label{t:extraps}
        \end{table}
         }        

We begin by extracting the linear functions $p(n)$ and $q(n)$ from the rational function $f(n)$ defining an endpoint family. In a first case, we have $q/p\equiv 0\mod 1.$ The action in~\eqref{eq:orbAction} is then trivial when $p=\pm1$. Here we extrapolate to $\emptyset,$ the empty endpoint. Other values of $p$ do no occur in this case.

Next, we consider the case with $p,q>0,$ for which the endpoint can be obtained as usual via~\eqref{eq:contFrac}. If on the other hand, $p<0,$ we reverse both the resulting extrapolated endpoint (for consistency with~\ref{eq:contFrac} as in the positive case) and the sign of $q/p.$ This allows endpoint extrapolation consistent with~\eqref{eq:orbAction} and endpoint reversal obtained via $\overline{f}(n)$ (even when these have opposite signs) via the rational $-q/p\mod1$. 

The final case is $p>0.$ Define $r:=q/p\mod1.$ We arrive at an endpoint defined by $r$ using~\ref{eq:contFrac} as usual.

A few notes are merited concerning the $T^*$ duality that appears via level reflection in the tower of endpoints. Briefly, the result is that a perfect mirror of negative level endpoints appears as the natural truncation consistent level inversion. A lattice of endpoints isomorphic to $\Z^2$ formed by endpoint extrapolations that precisely agrees with the partial order on endpoints inherited from canonical base truncations. The natural level inversion appears in this infinite grid of endpoints extrapolations as transposition about the antidiagonal where the only singular values of $f(n)$ arise to prevent extrapolations. This pairs identical endpoints arising via family swaps precisely according to the $T^*$ duality. This interchange appears at a half integer value of $n,$ a perhaps superficially curious fact at first glance. This turns out to be an obvious requirement for consistency with truncation partial order but appears unnatural in a table presentation of endpoints extrapolated from $f(n)$ since for example moving along the diagonal of symmetric endpoints results in level drops by steps of two. In other words, $T^*$ duality {\it is} level inversion.

\end{section}

\begin{section}{Expression of discrete $U(2)$ subgroup generators from leads/tails}\label{s:contFrac}
We pause to review a condensed description of the fraction formulas in Table~\ref{t:nFracs} taken from~\cite{gsThesis}. This identity illustrates that certain patterns appear among $f(n)$ that are related to lead/tail continued fractions but also depend critically on their matrix representations. This makes clear why certain superficial patterns are broken in some cases, namely those where nontrivial terms in these matrix representations come into play.

It will be convenient to first collect values for the continued fractions for leads/tails. These appear in Table~\ref{t:continuedFractionValues}. For convenience, we write $p_\alpha ,q_\alpha$ in place of $p(\alpha),q(\alpha),$ respectively, except for the empty lead $\alpha\sim \emptyset$ where we write $p_\alpha=0.$
First column entries of the Hirzebruch-Jung continued fraction matrix representation~\cite{Reidsurface} for the leading string, $\alpha,$ play a key role. Negations of these values in Table~\ref{t:auxFractionValues} which we label as $m_{\alpha,1},m_{\alpha,2}.$  
{\tiny \selectfont \renewcommand{\arraystretch}{1.8} \setlength{\tabcolsep}{5pt}{
\begin{table}[htbp]
\begin{center} 
\begin{tabular}[m]{|c|c|c|c|c|c|c|c|c|c|c|c|c|c|c|c|c|c|c|}
\hline 
$\alpha:$ & $\emptyset$ & 3 & 4 & 5 & 6 & 7 & 33 & 23 & 32 & 223 & 322 & 2223 & 3222 & 22223 & 32222 & 24 & 42   \\
\hline
$\frac{p_\alpha}{q_\alpha}:$ & 0 & 3 & 4 & 5 & 6 & 7 & $\frac{8}{3}$ & $\frac{5}{3}$ & $\frac{5}{2}$ & $\frac{7}{5}$ & $\frac{7}{3}$ & $\frac{9}{7}$ & $\frac{9}{4}$ & $\frac{11}{9}$ & $\frac{11}{5}$ & $\frac{7}{4}$ & $\frac{7}{2}$   \\
\hline
\end{tabular}
  \caption{Continued fraction values for lead endpoints.}
 \label{t:continuedFractionValues} 
\end{center}
\end{table}}}

\begin{table}[htbp]
\begin{center}
\begin{tabular}[m]{|c|c|c|c|c|c|c|c|c|c|c|c|c|c|c|c|c|c|}
\hline
$\alpha:$  & 3,4,5,6,7 & 23 &  223 & 2223 & 22223  & 33 & 24   \\
\hline
$(m_{\alpha,1} \ m_{\alpha,2}):$ & $(0 \ 1)$  & $(1 \ 2)$ & $(2 \ 3)$ & $(3 \ 4)$ & $(4 \ 5)$ & $(1\ 3)$ & $(1 \ 2)$  \\
\hline
\end{tabular} 
 \caption{First column entries of the Hirzebruch-Jung continued fraction matrix representation for endpoint strings.}
 \label{t:auxFractionValues}
\end{center}
\end{table}
The 78 permitted 6D SCFT paired $f(n)$ giving all orbifold isomorphism classes for endpoints of the form $\alpha A_n \beta$ with $\alpha,\overline{\beta}\in H$ can then be expressed as

\begin{align} \label{eq:fractionIdentity} 
\frac{p}{q}(\alpha A_n \beta) = 
\dfrac{ |(p_{\bar{\alpha}} - q_{\bar{\alpha}})(p_{\beta}-q_{\beta})| \cdot n
+|\max\{p_\alpha,1\}  \max\{p_\beta,1\} - q_\beta m_{\alpha,2}(1-\delta_{\beta,\emptyset} )  |  }         
{|(q_{\alpha}- m_{\alpha,1}) (p_{\beta} -q_{\beta})|
\cdot n
+ |q_\alpha \max\{p_\beta,1\} - q_\beta m_{\alpha,1} (1-\delta_{\beta,\emptyset} )| }.
\end{align}  

\noindent The absolute values, $\max(-),$ and Kronecker symbol are included so that we may simultaneously treat the cases with empty strings $\alpha,\beta$ and may otherwise be ignored. Absolute values are in some cases only included to emphasize that these $f(n)$ contain only nonnegative integer coefficients.
\end{section}

\begin{section}{Miscellaneous tables, D-type endpoints, more on endpoint combinatorics}\label{s:misc}
Here we extend our discussion of the rational functions determining linear 6D SCFT endpoints to those of D-type. We also collect tables peripheral to our main discussion and discuss further observations on endpoint combinatorics. In particular, we outline consequences of the observation that the rational functions $f(n)$ can be viewed as projective linear fractional transformations in $PSL(2,p)$ for certain $p.$

{ \tiny
     \begin{table}[htbp] 
    \begin{center}
    {   \fontsize{6.5}{8.8}\selectfont   \setlength{\tabcolsep}{2pt}
      \begin{tabular}{c|cccccccccccc}\diaghead(1,-1){cccccccc}{$\ \alpha$}{$\beta \ $}& 3222 & 33 & 42 & 6 & 322 & 5 & 4 & 32222 & 3 & 32  & $\emptyset$ \\
      \hline  \\
$\underset{22223}{}$ & $\underset{22223A_n 3222}{\overset{6A_m7}{\updownarrow}} $ & $\underset{22223A_n 33}{\overset{24A_m7}{\updownarrow}} $ & $ \underset{22223A_n 42}{\overset{33A_m7}{\updownarrow}} $ & $ \underset{22223A_n6}{\overset{2223A_m7}{\updownarrow}} $ & $ \underset{22223A_n 322}{\overset{5A_m7}{\updownarrow}} $ & $ \underset{22223A_n 5}{\overset{223A_m 7}{\updownarrow}} $ & $ \underset{22223A_n 4}{\overset{23A_m 7}{\updownarrow}} $ & $ \underset{22223A_n 32222}{\overset{7A_m 7}{\updownarrow}} $ &  
$ \underset{22223A_n 3}{\overset{3A_m 7}{\updownarrow}} $ &
$ \underset{22223A_n 32}{\overset{4A_m 7}{\updownarrow}} $ &
$ \underset{22223A_n \emptyset}{\overset{\emptyset A_m 7}{\updownarrow}} $ \\ \\
$\underset{6}{}$ & $\dagger$ &  $ \underset{6A_n 33}{\overset{24A_m 3222}{\updownarrow}} $ &
$ \underset{6A_n 42}{\overset{33A_m 3222}{\updownarrow}} $ &
$ \underset{6A_n6}{\overset{2223A_m 3222}{\updownarrow}} $ &
$ \underset{6A_n 322}{\overset{5A_m 3222}{\updownarrow}} $ &
$ \underset{6A_n 5}{\overset{223A_m 3222}{\updownarrow}} $ &
$ \underset{6A_n 4}{\overset{23A_m 3222}{\updownarrow}} $ & $*$ & 
$ \underset{6A_n 3}{\overset{3A_m 3222}{\updownarrow}} $ &
$ \underset{6A_n 32}{\overset{4A_m 3222}{\updownarrow}} $ &
$ \underset{6A_n \emptyset^{\not \ddagger}}{\overset{\emptyset A_m 3222^\ddagger}{\updownarrow}} $ \\ \\
$\underset{24}{}$ & $**$ & $\dagger$ &  $ \underset{24A_n 42}{\overset{33A_m 33}{\updownarrow}} $ & $*$ & 
$ \underset{24A_n 322}{\overset{5A_m 33}{\updownarrow}} $ &
$ \underset{24A_n 5}{\overset{223A_m 33}{\updownarrow}} $ &
$ \underset{24A_n 4}{\overset{23A_m 33}{\updownarrow}} $ & $*$ & 
$ \underset{24A_n 3}{\overset{3A_m 33}{\updownarrow}} $ &
$ \underset{24A_n 32}{\overset{4A_m 33}{\updownarrow}} $ &
$ \underset{24A_n \emptyset^{\not \ddagger}}{\overset{\emptyset A_m 33^\ddagger}{\updownarrow}} $ \\ \\
$\underset{5}{}$ & $**$ & $**$ & $*$ & $*$ & $\dagger$ & $ \underset{5A_n 5}{\overset{223A_m 322}{\updownarrow}} $ &
$ \underset{5A_n 4}{\overset{23A_m 322}{\updownarrow}} $ & $*$ &  
$ \underset{5A_n 3}{\overset{3A_m 322}{\updownarrow}} $ &
$ \underset{5A_n 32}{\overset{4A_m 322}{\updownarrow}} $ &
$ \underset{5A_n \emptyset^\ddagger}{\overset{\emptyset A_m 322^\ddagger}{\updownarrow}} $ \\ \\
$\underset{23}{}$  & $*$ & $**$ & $*$ &  $*$ & $**$ & $*$& $\dagger$  & $*$ & $ \underset{23A_n 3}{\overset{3A_m 4}{\updownarrow}} $ &
$ \underset{23A_n 32}{\overset{4A_m 4}{\updownarrow}} $ &
$ \underset{23A_n \emptyset^\ddagger}{\overset{\emptyset A_m 4^\ddagger}{\updownarrow}} $  \\
     \end{tabular}
     \\
     \vspace{0.5cm}
     
     \tiny \setlength{\tabcolsep}{7pt}
           \begin{tabular}{cccccccc} \multicolumn{8}{c}{\text{\scriptsize Self-dual endpoint families}} \\
           \hline \\
           $22223A_n 7$ & $2223A_n6$ & $223 A_n 5$ & $23A_n 4$ & $3A_n 3^\ddagger$ & $3 A_n \emptyset^\ddagger$ & $\emptyset A_n \emptyset^\ddagger$ & $24A_n33$ 
           \end{tabular}
      }
        \caption{Endpoint family pairings. Those families corresponding to $*$ and $**$ entries appear elsewhere in the table in lower and upper positions, respectively. Those indicated with `$\dagger$' are self-dual. The $\ddagger$ and $\not\text{\hspace{-0.05in}}\ddagger$ symbols indicate infinitely many permitted gauge enhancements for $n=0,$ and $n=-1$ or $n=-2,$ respectively.}
        \label{t:thePairing}
           \end{center}
        \end{table}
 }

\begin{subsection}{D-type rational functions $f(n).$}\label{s:dTypes}
In place of~\eqref{eq:fracDetRelation}, the D-type endpoint family rational functions obey
\begin{align}\label{eq:fracDetRelationDTypes}
-2\det(f(n)) = \gcd(k_u,k_l)^2 \ .
\end{align}
Note the additional factor of $2$ versus~\eqref{eq:fracDetRelation}. After minimal $n$-shift from the forms dating to~\cite{classifyingSCFTs}, these appear as
\begin{align}\label{eq:dTypeFracs}{\renewcommand*{\arraystretch}{2.3}
\begin{array}{cc}
D_{\widetilde{n}} 24 \leftrightarrow \dfrac{18n +6}{6n+1} \ , \qquad \qquad&
D_{\widetilde{n}} 32 \leftrightarrow \dfrac{18n +3}{12n+1} \ ,\\
D_{\widetilde{n}} 23 \leftrightarrow \dfrac{8n +4}{4n+1} \ , \qquad \qquad& 
D_{\widetilde{n}} 22 \leftrightarrow \dfrac{2n +2}{2n+1} \ ,
\end{array}}
\end{align} 
where $\widetilde{n}$ is an $n$-shift $D_N$ is the arrangement of $-2$ curves while $D_N \alpha$ for $\alpha = x_1\cdots x_n$ denotes
\begin{align}2\underset{N-2}{\underbrace{\overset{2}{2} 2 \cdots 2}} \alpha \ .
\end{align}
\noindent One can readily confirm that D-type endpoints are related to their defining $f(n)$ forms in a manner consistent with the row expressions of fixed $k_u/k_l$ for other endpoints.

Recall from~\cite{classifyingSCFTs} the Hirzebruch-Jung continued fraction for D-type $\Gamma$ generator can be obtained from the pairing
\begin{align}\label{eq:dTypeForm}
2 \overset{2}{y}m_1 m_2 \cdots m_l \quad \leftrightarrow \quad m_l \cdots m_2 m_1 (2y-2) m_1 m_2 \cdots m_l \ ,
\end{align}
which yields the expressions in~\eqref{eq:dTypeFracs} up to $n$-shift via
\begin{align}\label{eq:dTypeToAType}{\renewcommand*{\arraystretch}{2.3}
\begin{array}{cc}
D_{\widetilde{n}} 24\ \  \leftrightarrow  \ \ 4 A_{2n+1} 4, \qquad \qquad&
D_{\widetilde{n}} 32\ \  \leftrightarrow  \ \ 23 A_{2n+1} 32,\\
D_{\widetilde{n}} 23\ \  \leftrightarrow  \ \ 3 A_{2n+1} 3, \qquad \qquad& 
D_{\widetilde{n}} 22\ \  \leftrightarrow  \ \   A_{2n+1}\ .
\end{array}}
\end{align} 
These are hence manifestly endpoint-orientation independent. We can naturally place entries from~\eqref{eq:dTypeFracs} in a new column of our $f(n)$ table based upon their $k_u/k_l$ ratios.

For linear endpoints it is sufficient to determine the permitted fraction formulas using~\eqref{eq:fracDetRelation} and fixing allowed large $N$ limit values. The same does not hold for D-types where these constraints leave $11$ rather than $4$ permitted $f(n).$ The extra cases include $f(n)$ arising from $\alpha A_{2n} \beta$ having designated $(k_u,k_l).$ More generally, positive integer rescaling the left hand side of~\eqref{eq:fracDetRelation} and restricting $(k_u,k_l)$ to valid large $N$ limits simply yields $f(n)$ corresponding to subfamilies of endpoints expressed elsewhere in the table. Cases corresponding to D-types are simply those having an additional interpretation when a second generator acting by $(z_1,z_2)\mapsto (z_2, -z_1)$ yields non-isomorphic orbifolds. In this sense, these $p/q$ values are captured already by the determinant characterization above while a secondary structure gives them further meaning. Note that there are three exceptional ($\e_6,\e_7,\e_8$) branching endpoints which have not been treated in our discussion.

\end{subsection}

\begin{subsection}{SCFT endpoint families and $PSL(2,p)$}
Homomorphisms from the finite ADE subgroups $\Gamma\subset \su(2)$ into $E_8$ have been studied at length via geometric 6D SCFT constructions that have provided an alternate route to the classification of these homomorphisms from the mathematics literature~\cite{atomic,rgFission,heckman2019top}. Similarly, homomorphisms $SL(2,5)\to E_8$ have also been classified via geometric tools using 6D SCFT F-theory models~\cite{frey20186d}, consequently revealing errors in their extant classification from the mathematics literature while also enabling a novel result: the classification of homomorphisms from the binary dihedral, tetrahedral, and octahedral groups into $E_8$. Discrete $U(2)$ gauge fields associated to 6D SCFTs in contrast do not exhaust the list of all finite $U(2)$ subgroups. Those that pair with 6D SCFTs remain to be completely matched to a known mathematical structure. This situation motivates our work here to better characterize endpoint landscape combinatorics, including the Section~\ref{s:dets} discussion relating SCFT fraction determinants and the intermediate determinants $\Gamma\to U(1).$ In the same direction, we make the following observation that a homomorphism from a cyclic group into $E_8\times E_8$ arises from each linear endpoint family.

To construct these homomorphisms, we consider the behavior of $\det(f(n))$ for $f(n)$ defining a valid endpoint family. Since the fraction determinant of $1/f(n)$ is a square, reduction modulo certain primes $p$ yields $[1/f(n)],$ which can be viewed as a linear fractional transformation over the finite field $\F_p.$ (Alternatively, we can reverse level indexing with the replacement $n\mapsto -n$.) For typical $p$, the resulting permitted values of $\det(f(n))$ are squares modulo $p.$ For $p\leq 13$, these exhaust the squares in $\F_{p}.$ This associates to each $[1/f_{i,j}(n)]$ an element of $G=PSL(2,p)$ for various various choices of prime $p\geq 7.$ Only for symmetric endpoints is the result independent of endpoint reversal. Truncation induced endpoint ordering naturally associates the collection of linear endpoints with a subset in $PSL(2,p)$ (and in a distinct way in $PSL(2,p)\times PSL(2,p)$). Since these groups naturally embed in $E_8$ for certain $p,$ via the classification~\cite{griess} of simple group homomorphisms into $E_8,$ we can associate to each endpoint family an embedding of a cyclic group into $E_8.$ The pair $f_{i,j},f_{j,i}$ similarly defines a map from another cyclic group into $E_8\times E_8$. The smallest $p$ for which the latter map determines unique pairs of elements in $PSL(2,p)$ corresponding to endpoint families is 11. Considering all level shift and reversal compatible $f(n)$ associated to endpoint families in this way naturally leads us to collapse the collection of linear 6D SCFT paired $f(n)$ with $PSL(2,p),$ even if perhaps only as a set. 

This makes it somewhat notable that the above maps factor through certain simple groups for some $p.$ For example, via Carmichael's constructions~\cite{carmichael1937introduction} of the Steiner systems $S(5,6,12)$ and $S(5,8,24)$ with automorphism groups $M_{12},M_{24},$ we see that each $f(n)$ can be viewed in each group by its action on blocks. Recall that this approach involves letting linear fractional transformations act on the blocks, for example with $S(5,6,12)$ the block of squares in $P_{11}$ 
\begin{align}\label{eq:seed}
B\sim \{ \infty, 1,3,4,5,9\}
\end{align}
gives under these actions
\begin{align}
z \mapsto \frac{a z + b}{c z + d} \qquad \text{ such that } \qquad a,b,c,d \in P_{11}\ , \ \ \det(f(z)) = \overline{s^2}\in P_{11}
\end{align}
the blocks of $S(5,6,12).$ In other words, we can view each oriented SCFT endpoint family as an element (or its generated cycle) in $PSL(2,11)$ and consequently also as an element in $M_{11}.$
Similarly reducing modulo $23$ leads to elements in $PSL(2,23)$ and embeddings into $M_{24}$ via action instead on
\begin{align}\label{eq:seed2}
B\sim \{ \infty, 0, 1, 3, 12, 15, 21, 22\} \ 
\end{align}
as a block with elements in $P_{23}.$
\end{subsection}

\end{section}
\FloatBarrier

\FloatBarrier

\bibliographystyle{unsrt}

\end{document}